\documentclass[12pt]{article}
\usepackage{amsmath,amssymb,amsfonts,amsthm,latexsym,color,url}
\usepackage{graphicx,psfrag,epsf}
\usepackage[ruled]{algorithm2e}
\usepackage{graphics}
\usepackage{geometry}
\usepackage{lscape}
\usepackage[font=normalsize]{caption}
\usepackage{scalerel}
\usepackage[utf8]{inputenc}
\usepackage[english]{babel}
\usepackage[titletoc]{appendix}
\usepackage{enumitem}
\usepackage{authblk}
\usepackage{multirow}
\usepackage{tabularx}
\usepackage{natbib}
\usepackage{url} 
\usepackage{pdfpages}

\newcommand{\blind}{0}

\def\sI{{\mathcal I}}                            

\def\sS{{\mathcal S}}
\def\sP{{\mathcal P}}

\def\sB{{\mathcal B}}
\def\sD{{\mathcal D}}

\def\sE{{\mathcal E}}

\def\sR{{\mathcal R}}

\def\sA{{\mathcal A}}
\def\sZ{{\mathcal Z}}

\def\sK{{\mathcal K}}
\def\sG{{\mathcal G}}

\def\vectorfontone{\bf}
\def\vectorfonttwo{\boldsymbol}
\def\vb{{\vectorfontone b}}                      %
\def\vc{{\vectorfontone c}}                      %
\def\vp{{\vectorfontone p}}                      
\def\vu{{\vectorfontone u}}                      
\def\vx{{\vectorfontone x}}                      
\def\vy{{\vectorfontone y}}                      
\def\vz{{\vectorfontone z}}                      %

\def\vone{{\vectorfontone 1}}
\def\vzero{{\vectorfontone 0}}

\def\vbeta{{\vectorfonttwo \beta}}               
\def\vgamma{{\vectorfonttwo \gamma}}             %
\def\vepsilon{{\vectorfonttwo \epsilon}}         %
\def\vtheta{{\vectorfonttwo \theta}}             
\def\vmu{{\vectorfonttwo \mu}}                   
\def\matrixfontone{\bf}
\def\matrixfonttwo{\boldsymbol}
\def\mB{{\matrixfontone B}}                      %
\def\mD{{\matrixfontone D}}                      
\def\mI{{\matrixfontone I}}                      
\def\mJ{{\matrixfontone J}}                      %
\def\mM{{\matrixfontone M}}                      %
\def\mU{{\matrixfontone U}}                      
\def\mV{{\matrixfontone V}}                      %
\def\mW{{\matrixfontone W}}                      
\def\mX{{\matrixfontone X}}                      
\def\mY{{\matrixfontone Y}}                      %

\def\mDelta{{\matrixfonttwo \Delta}}             %
\def\mSigma{{\matrixfonttwo \Sigma}}             %
\def\mPsi{{\matrixfonttwo \Psi}}                 %

\def\bE{{\mathbb E}}                             
\def\bP{{\mathbb P}}                             
\def\bR{{\mathbb R}}                             

\def\argmax{\operatornamewithlimits{\text{argmax}}}

\def\tr{\mbox{tr}}

\def\argmin{\operatornamewithlimits{\text{argmin}}}

\newtheorem{theorem}{Theorem}
\newtheorem{lemma}{Lemma}

\definecolor{orange}{rgb}{1,0.5,0}
\definecolor{purple}{rgb}{0.75,0,1}
\definecolor{darkgreen}{rgb}{0,0.5,0}
\definecolor{brown}{rgb}{0.59, 0.29, 0.0}

\newlength{\bibitemsep}
\newlength{\bibparskip}
\let\oldthebibliography\thebibliography
\renewcommand\thebibliography[1]{%
	\oldthebibliography{#1}%
	\setlength{\parskip}{\bibitemsep}%
	\setlength{\itemsep}{\bibparskip}%
}

\renewcommand{\baselinestretch}{1.5} 

\begin{document}
    
\title{Variational approximate penalized credible regions for Bayesian grouped regression}

\author{Weichang Yu\thanks{
			This work is partially funded by the \textit{Australian Research Council}. } \thanks{School of Mathematics and Statistics,  The University of Melbourne, Australia} and 
		Khue-Dung Dang \thanks{School of Physics, Mathematics and Computing, The University of Western Australia, Australia} \\
\date{}        
		}
\maketitle

\begin{abstract}
We develop a fast and accurate grouped penalized credible region approach for variable selection and prediction in Bayesian high-dimensional regression. Most existing Bayesian methods either are subject to high computational costs due to long Markov Chain Monte Carlo runs or yield ambiguous variable selection results due to non-sparse solution output. The penalized credible region framework yields sparse post-processed estimates that facilitates unambiguous grouped variable selection. High estimation accuracy is achieved by shrinking noise from unimportant groups using a grouped global-local shrinkage prior. To ensure computational scalability, we approximate posterior summaries using coordinate ascent variational inference and recast the penalized credible region framework as a convex optimization problem that admits efficient computations. We prove that the resultant post-processed estimators are both parameter-consistent and variable selection consistent in high-dimensional settings. Theory is developed to justify running the coordinate ascent algorithm for at least two cycles. Through extensive simulations, we demonstrate that our proposed method outperforms state-of-the-art methods in grouped variable selection, prediction, and computation time for several common models including ANOVA and nonparametric varying coefficient models.
\end{abstract}

\noindent%

\section{Introduction}
\label{sec::Introduction}
\noindent Consider a regression model for the dataset $\sD = \{(\breve{y}_i, \vz_i) \}_{i=1}^n$, where each $\vz_i \in \sZ$ is a  vector of covariates and $\breve{y}_i \in \bR$ is a univariate response:
\begin{equation}
	\label{eqn::GeneralRegression}
	\breve{y}_i = f_{\vbeta}(\vz_i) + \breve{\epsilon}_i,
\end{equation}
and $\breve{\epsilon}_i \sim N(0, \sigma^2)$. In several scenarios, we can justify a decomposition of the regression function as an intercept plus sum of dot products
\begin{equation}
	\label{eqn::dotproductdecomposs}
	f_{\vbeta}(\vz_i) = \kappa + \sum_{g=1}^G \widetilde{\vx}_g(\vz_i)^\top \vbeta_g.
\end{equation}
Such decompositions may be motivated by application-specific considerations. For example, in gene expression analysis, the vector $\widetilde{\vx}_g(\vz_i)$ encodes gene expression levels along the same biological pathway. Other justifications arise intrinsically from modeling assumptions. For example, in additive splines regression, $\widetilde{\vx}_g(\vz_i)$ represents a splines basis for the $g$-th covariate. Other examples include multi-way ANOVA models and varying coefficient models \citep{Hastie1993}. In these settings, the decomposition naturally motivates a grouped variable selection framework, where variables are selected at the group level rather than individually.

Most existing methods for grouped variable selection are developed within the frequentist paradigm \citep{yuan2006model, Bach2008, Wang2008b, Zhao2009, Huang2012, Geng2015, Beer2019, Thompson2024}. Bayesian approaches to grouped variable selection instead rely on assigning suitable priors to regression coefficients to encourage group-wise sparsity \citep{Xu2016,Tang2018, Yang2020}. Broadly, such priors can be classified into two categories as extensions of their non-grouped counterparts: \textit{two-component priors} and \textit{one-component priors}. Despite their conceptual appeal, many methods based on these priors do not directly quantify group importance or may incur substantial computational costs.

A well-known example of a two-component prior is the spike-and-slab prior \citep{Mitchell1988}, where each coefficient is assigned an independent mixture of a discrete point mass at zero (spike) and a continuous distribution with large variance (slab). A popular formulation of the spike-and-slab prior, known as \textit{stochastic search variable selection} (SSVS), was proposed in \cite{McCulloch1993}, where each coefficient group (or individual coefficient) is assigned with a binary latent indicator to indicate a presence or absence of effect on the response variable. The importance of each variable group (or individual variable) can be inferred directly from the posterior of the latent indicators. Extensions of SSVS to grouped variable selection have been studied in \cite{Farcomeni2010}, \cite{Xu2015}, \cite{Chen2016}, and \cite{Yang2020}. While an advantage of the spike-and-slab prior is that it facilitates direct inference of each group's importance, posterior inference over a large number of binary indicators can be computationally prohibitive, especially in high-dimensional problems. An existing computational trick is to marginalise over the binary indicators and report a sparse posterior summary statistic \citep{Xu2015}. To reduce computational costs, \cite{Rockova2018} proposed the spike-and-slab LASSO which replaces the discrete point masses with a continuous distribution and posterior inference of the binary indicators is avoided through marginalisation. The spike-and-slab LASSO prior is adapted for grouped-variable selection in \cite{Bai2022}. In fact, to identify significant groups they propose using the posterior mode which they prove to be sparse.

A common subclass of one-component priors is the \textit{global-local shrinkage prior} \citep{Polson2011}. This subclass of priors for the coefficient groups (or individual coefficients) are expressed as a scale mixture of Gaussians, where each prior variance is a product of a global shrinkage parameter and a local variable-specific parameter that potentially counteracts the shrinkage effect from the global parameter. Examples in the non-grouped setting include the Strawderman-Berger prior \citep{Strawderman1971,Berger1980}, the t-prior \citep{Tipping2001}, the Bayesian Lasso \citep{Park2008}, the normal-gamma prior \citep{Griffin2010}, the Horseshoe and Horseshoe$+$ priors \citep{Carvalho2010, Bhadra2017}, and the Dirichlet-Laplace prior \citep{Bhattacharya2015}. Examples of global-local-shrinkage priors for the grouped variable selection include the Bayesian grouped LASSO prior \citep{Kyung2010} and the grouped horseshoe prior \citep{Xu2016}. Beyond global-local shrinkage priors, other examples of grouped variable selection priors include the grouped R2-D2 prior \citep{Yanchenko2024}. Many examples of one-component prior methods require MCMC sampling for posterior computations. However, the resulting MCMC sampler is susceptible to poor mixing, often requiring long runs to obtain posterior samples that adequately explore the posterior space. Moreover, posterior expectations based on one-component priors are not strictly sparse thus prohibiting variable selection. To address this gap, \cite{Paul2025} propose a norm-based thresholding rule to post-process the group-specific posterior expectations.

Another promising alternative approach to post-process non-sparse posterior summaries is the penalized credible region \citep{Bondell2012}. In this approach, the posterior expectation and covariance of all coefficients under the full model is obtained. Then, a sequence of Bayesian credible regions is constructed such that it corresponds to an increasing sequence of credibility levels. The corresponding sequence of sparsest solution is obtained as an estimate of the target coefficient vector, thus aligning with the Occam's razor principle. Theoretical guarantees for penalized credible region estimates based on the normal prior and the Dirichlet-Laplace prior have been established for non-grouped variable selection \citep{Bondell2012, Zhang2018}.

In this paper, we propose a fast and accurate Bayesian grouped variable selection procedure by combining penalized credible region with a one-component prior. Unlike the original version of the procedure in \cite{Bondell2012} where a normal prior is assigned to enable exact posterior moment computations, we assign the coefficients with a grouped horseshoe prior, thus resulting in a \textit{shrink-and-sparsify} posterior estimator. Note that our choice of the grouped horseshoe prior is heuristically motivated by the theoretical guarantees and excellent numerical results of its non-grouped analogue \citep{vanderPas2017}. Here, we encounter a computational issue with the intractable posterior moments. To enable fast computations, we approximate the posterior moments through coordinate-ascent variational inference (CAVI) \citep{Ormerod2010, Blei2017}. In fact, CAVI has shown to enable extremely fast approximate posterior computations in several examples. We prove both parameter and variable selection consistencies of our proposed estimator and provide theoretical insights about the use of CAVI in our proposed procedure.


Our main contributions are threefold: (i) To the best of our knowledge, our proposed method is the first penalized credible region for grouped variable selection method that enforces group-level sparsity; (ii) We provide thorough justifications of our grouped horseshoe prior through excellent numerical results and prove both the parameter estimation consistency and variable selection consistency. On the other hand, preceding research on grouped horseshoe priors by \cite{Xu2016} focuses on its performance in parameter estimation and not variable selection. Moreover, no asymptotic properties was discussed; (iii) Prior penalized credible region methods rely on exact posterior moment calculations or Gibbs sampling \citep{Bondell2012, Zhang2018}. We instead employ CAVI to approximate posterior moments, which raises fundamental questions regarding the validity of combining penalized credible region sparsification with variational approximations and the number of CAVI iterations to run. These issues are explicitly addressed in our analysis. From a theoretical perspective, most existing work on variational Bayes establishes posterior concentration or asymptotic normality of the variational posterior itself \citep{Lu2017, You2017, Wang2019, AlQuier2020, Zhang2020c, Ray2022}, while \cite{Mukherjee2022} focuses on asymptotic accuracy in high-dimensional linear regression under a product prior setup. In contrast, we study the consistency properties of post-processed variational posterior summaries, proving both parameter estimation consistency and variable selection consistency. Furthermore, in contrast to existing works \citep{Wang2006, Zhang2020b} that
provides theoretical guarantees on the estimation accuracy improvement of variational parameters
with increasing number of CAVI iterations, we provide theoretical guarantees for estimation accuracy
improvement of our post-processed variational Bayes estimator with increasing number of
CAVI iterations.

The remainder of the paper is organized as follows. Section \ref{sec::ProblemSetup} reviews grouped variable selection, introduces the grouped horseshoe prior, and presents the grouped penalized credible region framework. Section \ref{sec::CAVIreview} describes the CAVI algorithm and the proposed variational penalized credible region procedure. Section \ref{sec::theory} provides a theoretical justification of our proposed framework to address important questions about the usage of our method. Section \ref{sec::numericalStudy} compares the numerical performance of our proposed method with state-of-the-art alternatives in both simulated and real datasets. Section \ref{sec::ConclusionFutureWork} concludes.

\section{Problem setup}
\label{sec::ProblemSetup}
We focus on the cases where the decomposition \eqref{eqn::dotproductdecomposs} is justified. Then, the regression model in \eqref{eqn::GeneralRegression} may be expressed as:
\begin{equation}
	\label{eqn::UncenteredModel}
	\breve{y}_i = \kappa + \sum_{g=1}^G \breve{\vx}_{ig}^\top \vbeta_g + \breve{\epsilon}_i,
\end{equation}
where $\breve{\vx}_{ig} := \widetilde{\vx}_g (\vz_i)$, $\vbeta_g = (\beta_{g1}, \ldots, \beta_{g p_g} )$ is the $g$-th subvector of regression coefficients, $\kappa \in \bR$, and $\breve{\epsilon}_i \sim N(0, \sigma^2)$. Assume that the true relationship between the response and the covariates is $\breve{y}_i = \kappa^0 + \sum_{g \in \sA^0} \breve{\vx}_{ig}^\top \vbeta_g^0  + \epsilon_i^\star$, and $\epsilon_i^\star \sim N(0, \sigma^{\star \, 2})$, where $\sA^0 \subseteq [G] := \{ 1, \ldots, G\}$ and $\vbeta_g^0 \neq \vzero$. Our main objectives are to identify the \emph{true set} $\sA^0$, develop an accurate Bayesian point estimate of $\vbeta^0$, and predict unobserved responses. In many applications of high-dimensional regression, the true coefficient vector $\vbeta^{0} = ( \vbeta_1^{0 \, \top}, \ldots, \vbeta_G^{0 \, \top} )^\top$ is assumed to be sparse. Likewise, we assume that $\vbeta^0$ is sparse; we discuss the details in section \ref{sec::theory}. Prior assignment for inference of sparse coefficient vectors is non-trivial. If our prior is too flat, we may end up with many spurious signals from non-significant groups. On the other hand, if our prior places excessively-heavy mass around $0$, we may introduce excessive bias in the posterior point estimates.

\subsection{Grouped horseshoe prior}
To amplify group-shrinkage effect on the non-significant groups and reduce bias in posterior point estimates, we require a suitable grouped variable selection analogue of the global-local shrinkage prior. In the original non-grouped formulation of the global-local shrinkage prior with $p_g = 1$, the global local shrinkage prior can be expressed as:
\begin{align*}
	&\beta_g \mid \sigma^2, \tau, b_g \sim N(0, \sigma^2 \tau / b_g), \;\; b_g \sim p(b_g) \; \; \forall \; \; g = 1,\ldots, G, \\ 
	&(\sigma^2, \tau) \sim p(\sigma^2, \tau).
\end{align*}
Here, in our grouped variable selection setting, we assign the grouped horseshoe prior \citep{Xu2016} which has the form:
\begin{align}
	\label{eqn::groupedHorseshoe}
	&\vbeta_g \mid \sigma^2, \tau, b_g \sim N(\vzero, \tau \sigma^2 / b_g \mI_{p_g}), \; b_g  \sim \text{Ga}(1/2, c_g), \;  c_g  \sim \text{Ga}(1/2, 1) \; \; \forall \; \; g = 1,\ldots, G, \nonumber \\
	&\sigma^2 \sim \text{InvGa} (r, s),
\end{align}
where we treat $\tau$ as an optimization parameter and we adopt a hierarchical formulation of the horseshoe distribution as presented in \cite{Neville2014}. To simplify downstream computations, we avoid inferring $\kappa$ directly by subtracting $\overline{\breve{y}} = \tfrac{1}{n} \sum_{i=1}^n \breve{y}_i $ on both sides of (\ref{eqn::UncenteredModel}) to obtain a centered representation of our model:
\begin{equation*}
	\label{eqn::CenteredModel}
	y_i  = \sum_{g=1}^G \vx_{ig}^\top \vbeta_g + \epsilon_i,
\end{equation*}
where $\epsilon_i = \breve{\epsilon}_i - \overline{\breve{\epsilon}}$, $y_i = \breve{y}_i - \overline{\breve{y}}$ and $\vx_{ig} = \breve{\vx}_{ig} - \overline{\breve{\vx}}_{g \cdot}$.
Unlike the typical centered regression model, the centered error terms $\epsilon_i$'s are correlated. In particular, the true covariance matrix of the centered random errors is $\text{Var}( \vepsilon  ) = \sigma^{\star \, 2} \mPsi_n$, where $\mPsi_n = \mI_n - \tfrac{1}{n}\mJ_n$ and $\mJ_n$ is a $n$ by $n$ matrix of ones. However, we deliberately ignore the pairwise correlations between the centered errors in our posterior computations to reduce computational cost, though this correlation is fully accounted for in the asymptotic analysis in section \ref{sec::theory}. The joint posterior distribution of our specified model and auxiliary parameters is:
\begin{equation}
	\label{eqn::JointPostHorseshoe}
	p_\tau( \vbeta, \sigma^2, \vb, \vc \mid \sD) \propto \phi_n ( \vy; \mX \vbeta, \sigma^2 \mI) \times p( \vbeta \mid \sigma^2, \tau ) p( \sigma^2 ) p( \vb \mid \vc) p(\vc),
\end{equation}
where $\vb = (b_1, \ldots, b_G)$, $\vc = (c_1, \ldots, c_G)$, $\phi_n$ denotes the $n$-variate Gaussian density, and
$$
\mX = \begin{pmatrix}
	\vx_{11}^\top & \ldots & \vx_{1G}^\top \\
	\vdots & \vdots & \vdots \\
	\vx_{n1}^\top & \ldots & \vx_{nG}^\top
\end{pmatrix}.
$$
\noindent Following the joint posterior in (\ref{eqn::JointPostHorseshoe}), inference for the slopes proceeds from the marginal posterior
\begin{equation}
	p_\tau( \vbeta \mid \sD) = \int \int \int p_\tau( \vbeta, \sigma^2, \vb, \vc \mid \sD) \; \text{d} \sigma^2 \text{d} \vb \text{d} \vc.
\end{equation}
Since $p_\tau( \vbeta \mid \sD)$ is a density that is continuous everywhere in $\bR^p$, a na{\"i}ve approach to inferring the importance of each group $g$ using posterior probabilities yields $p_\tau (  \vbeta_g \neq \vzero \mid \sD) = 1$ for all $g \in [G]$ which is not useful. In fact, this problem is pervasive among most posteriors that are based on one-component priors.
\subsection{Group penalized credible regions}
We propose a group-penalized credible region approach to post-process posterior inference summaries. Note that while the final formulation of our proposed posterior estimate is based on a grouped horseshoe prior assignment, we describe the framework for any arbitrary prior choice $\vbeta \sim \pi(\vbeta)$. Denote the $100(1- \alpha)\%$ highest density region as $\sS_{\alpha; \pi} \subset \bR^p$, where $p = \sum_g p_g$. A na{\"i}ve adaptation of the original Bayesian penalized credible region formulation \citep{Bondell2012} is the sequence of solutions for $\widetilde{\vbeta}$ such that:
$$
\widetilde{\vbeta} = \argmin_{\vbeta \in \bR^{p}} \sum_{g=1}^G \sum_{j=1}^{p_g} I( \beta_{gj} \neq 0 ), \;\; \text{subject to} \;\; \vbeta \in \sS_{\alpha; \pi},
$$
where $\alpha$ is a decreasing sequence\footnote{In preceding papers \citep{Bondell2012, Zhang2018, Zhang2021BookChap}, the credible region $\sS_{\alpha, \pi}$ may be specified as any arbitrary region with posterior probability $1 - \alpha$. However, the resultant estimator is very sensitive to the choice of the credible region. To avoid ambiguity, we consider only the highest density region in our formulation.}.
Observe that the above formulation does not support grouped variable selection. More specifically, the resultant $\widetilde{\vbeta}_g$ may have some but not all entries equal to $0$, which leaves room for ambiguity when discerning group-level importance. To facilitate grouped variable selection, we propose the following formulation to enforce group-level sparsity:
\begin{equation}
	\label{groupedL0formulation}
	\widetilde{\vbeta} = \argmin_{\vbeta \in \bR^{p}} \sum_{g=1}^G \sqrt{p_g} I\{ u_g \neq 0 \}, \;\; \text{subject to} \;\; \vbeta \in \sS_{\alpha; \pi}.
\end{equation}
where $u_g :=  \lVert \vbeta_g \rVert $ and $\lVert \cdot \rVert$ denote the $L_2$-norm. Note that the $\sqrt{p_g}$ weights account for the differences in group sizes that may undesirably lead to higher selection rates for large groups in an unweighted objective function \citep{yuan2006model, Buhlmann2011}. Clearly, for a fixed $\pi$ and $\alpha$, there is no guarantee that the minimizer to (\ref{groupedL0formulation}) is unique which follows from the non-smooth indicator function. Consequently, the computation of $\widetilde{\vbeta}$ may be overly-complicated. To address this lack of smoothness, we replace the indicator-objective with a smooth homotopy:
$$
\rho_a (t) = \left ( \frac{t}{a+t} \right ) I( t \neq 0) + \left ( \frac{a}{a+t} \right ) t, \;\; t\ge 0, a>0.
$$
By a first-order local linear approximation argument on $\rho_a$ about $\widehat{u}_g  = \bE \{ \lVert \vbeta_g \rVert \mid \sD \}$ and then letting $a \rightarrow 0$, our optimization problem can be expressed as:
\begin{equation}
	\label{ConstraintForm}
	\widetilde{\vbeta} =\argmin_{\vbeta \in \bR^{p}}  \sum_{g=1}^G \sqrt{p_g} \frac{u_g}{ \widehat{u}_g^2  },  \;\; \text{subject to} \;\; \vbeta \in \sS_{\alpha; \pi}.
\end{equation}
\subsection{Computations}\label{subsec::Computations}

If $\sS_{\alpha; \pi}$ is a convex set, then the convexity of the optimization objective in \eqref{ConstraintForm} allows us to express $\widetilde{\vbeta}$ as a minimizer of a Lagrangian function. In particular, if $\sS_{\alpha; \pi}$ is elliptical, we have
\begin{equation}
	\label{DualForm}
	\widetilde{\vbeta} =\argmin_{\vbeta \in \bR^{p}} (\vbeta - \widehat{\vbeta})^\top \mSigma_{\vbeta}^{-1} (\vbeta - \widehat{\vbeta} ) + \lambda_\alpha \sum_{g=1}^G \sqrt{p_g} \frac{u_g}{ \widehat{u}_g^2  },
\end{equation}
where $\widehat{\vbeta}$ and $\mSigma_{\vbeta}$ are the posterior mean and covariance matrix. To obtain $\widetilde{\vbeta}$, we first solve a group-lasso regression with design matrix $\boldsymbol{X}^*$ and response vector $\mY^*$ as
$$\boldsymbol{X}^* = \mSigma_{\vbeta}^{-1/2}\boldsymbol{D}, \;  \mY^* = \mSigma_{\vbeta}^{-1/2}\widehat{\vbeta}, $$
where $\boldsymbol{D}$ is a block diagonal matrix with the $g^{th}$ diagonal element being the diagonal matrix with diagonal elements are $\widehat{u}_g^2$. The sequence of solution $\vbeta^*$ can be computed easily with existing algorithms for group-lasso. We implement the group-lasso regression algorithm using the \texttt{R} package \texttt{gglasso} \citep{yang2015fast} that is available on \texttt{CRAN}. Then $\widetilde{\vbeta} =  \boldsymbol{D}\vbeta^* $ is a solution to (\ref{ConstraintForm}). The justification for this approach to find $\widetilde{\vbeta}$ is provided in supplementary section A.

\noindent \textbf{Remark on computational challenges of grouped penalized credible regions}: While the formulation in \eqref{DualForm} facilitates efficient computation of $\widetilde{\vbeta}$, the formula is only applicable in cases where $\sS_{\alpha; \pi}$ is elliptical. For many specifications of $\pi$, including the global-local shrinkage prior, $\sS_{\alpha; \pi}$ is non-convex and consequently computation of the minimizer in \eqref{ConstraintForm} is not straightforward. To avoid working with a potentially non-convex credible region from a Dirichlet-Laplace prior specification, \cite{Zhang2018} restricted their choice of $\sS_{\alpha; \pi}$ to elliptical regions centered at $\bE\{ \vbeta \mid \sD \}$ and scaled by $\text{Var}\{ \vbeta \mid \sD \}$\footnote{A non-computational issue with choosing elliptical regions rather than the highest density region is that the credible regions may have overly large volume, thus leading to poor variable selection performance.}. While this solution works well in several examples and has good asymptotic properties, their method's computation requires a Gibbs sampler which is susceptible to poor mixing.  In fact, this issue is well-demonstrated in our numerical results in Section \ref{sec::numericalStudy}.

\subsection{Point prediction}
Our proposed posterior predictive distribution of a new uncentered response $\breve{y}^\star$ corresponding to an uncentered covariate $\breve{\vx}^\star$ is:
$$
p(\breve{y}^\star \mid \sD) = \int \phi_n( \breve{y}^\star; \kappa + \breve{\vx}^{\star \, \top} \vbeta, \sigma^2) p( \kappa \mid \vbeta, \sD) p_\tau( \vbeta, \sigma^2, \vb, \vc \mid \sD) \; \text{d}\vtheta,
$$
where $\vtheta = (\kappa, \vbeta, \sigma^2, \vb, \vc)$. To compute the posterior predictive, we require a specification for the conditional posterior $p( \kappa \mid \vbeta, \sD)$. Here, we specify the conditional prior of $\kappa$ as $\pi(\kappa \mid \vbeta) = \delta_{\widehat{\kappa} (\vbeta) } (\kappa)$ so that the induced conditional posterior is also of the form $p( \kappa \mid \vbeta, \sD) = \delta_{\widehat{\kappa} (\vbeta) } (\kappa)$, where $\delta_{a}$ denotes the Dirac delta measure on $a$. Then, we tune $\widehat{\kappa}$ to maximize the joint density between the observed data $\sD$ and $\vbeta$, i.e.,
\begin{align*}
	\widehat{\kappa} (\vbeta) &= \argmax_{\kappa} \log p( \sD, \vbeta \mid \kappa ) \\
	&= \argmax_{\kappa}  \int \int \int \phi_n \left ( \breve{\vy}; \kappa \vone_n + \breve{\mX} \vbeta, \sigma^2 \mI_n \right ) \prod_{g=1}^G \pi(\vbeta_g \mid \sigma^2, b_g) \pi(\sigma^2) \pi(\vb, \vc) \; \text{d} \sigma^2 \, \text{d} \vb \, \text{d} \vc \\
	&= \argmax_{\kappa}  \phi_n \left ( \breve{\vy}; \kappa \vone_n + \breve{\mX} \vbeta, \sigma^2 \mI_n \right ) \prod_{g=1}^G \pi(\vbeta_g \mid \sigma^2, b_g) \pi(\sigma^2) \pi(\vb, \vc), \\
	&= \argmin_{\kappa}   \lVert \breve{\vy} - \breve{\mX} \vbeta - \kappa\vone \rVert^2 \\
	&= \overline{\breve{y}} - \vbeta^\top \overline{\breve{\vx}},
\end{align*}
where the third equality follows by noting that the maximizer of the objective at the third equality does not depend on $\sigma^2$, $\vb$, or $\vc$. Here, $\overline{\breve{y}}$ and $\overline{\breve{\vx}}$ denote the sample response mean and sample predictor centroid respectively. Given a point in the predictor space $\breve{\vx}^\star$, a Bayes point prediction for $\breve{y}^\star$ is
\begin{align*}
	\bE \{ \breve{y}^\star \mid \sD \} &= \int \breve{y}^\star p(\breve{y}^\star \mid \sD) \; \text{d} \breve{y}^\star \\ &=  \overline{\breve{y}} + \left (\breve{\vx}^{\star} - \overline{\breve{\vx}} \right )^\top \bE \{ \vbeta \mid \sD \}.
\end{align*}
We replace $\bE \{ \vbeta \mid \sD \}$ with the grouped penalized credible region estimator $\widetilde{\vbeta}$ to obtain the proposed point prediction
$$
\widetilde{\mu} = \overline{\breve{y}} + \left (\breve{\vx}^{\star} - \overline{\breve{\vx}} \right )^\top \widetilde{\vbeta}.
$$

\section{Coordinate ascent variational inference}
\label{sec::CAVIreview}
To enable fast computations and accurate posterior estimation in our proposed method, we specify
$\sS_{\alpha; \pi}$ as the highest density region of an approximate posterior based on CAVI. In fact, CAVI belongs to the class of \textit{variational inference} methods for approximating an intractable joint posterior distribution such as (\ref{eqn::JointPostHorseshoe}). In CAVI, the joint posterior is approximated with a product of mean-field densities of the form:
\begin{align}
	\label{densityapprox}
	p_\tau( \vbeta, \sigma^2, \vb, \vc \mid \sD) &\approx q_\tau(\vbeta, \sigma^2, \vb, \vc) \nonumber \\
	&= q_\tau^{MF}(\vbeta) q_\tau^{MF}(\sigma^2) q_\tau^{MF}(\vb) q^{MF}(\vc)
\end{align}
where the density components on the RHS are chosen to minimize the KL-divergence between the target and the approximation or equivalently the evidence lower bound (ELBO), i.e.,
\begin{align*}
	q_\tau^{MF}(\vbeta) q_\tau^{MF}(\sigma^2) q_\tau^{MF}(\vb) q^{MF}(\vc) &=   \argmin_{\widetilde{q}} \text{KL} \left [ \widetilde{q} (\vbeta) \widetilde{q}(\sigma^2) \widetilde{q}(\vb) \widetilde{q}(\vc) \, || \, p_\tau( \vbeta, \sigma^2, \vb, \vc \mid \sD) \right] \\
	&= \argmax_{\widetilde{q}} \text{ELBO}(\widetilde{q})
\end{align*}
where
$$
\text{KL} \left [ \widetilde{q} (\vbeta) \widetilde{q}(\sigma^2) \widetilde{q}(\vb) \widetilde{q}(\vc) \, || \, p_\tau( \vbeta, \sigma^2, \vb, \vc \mid \sD) \right] = \bE_{\widetilde{q}} \left [ \frac{\widetilde{q} (\vbeta) \widetilde{q}(\sigma^2) \widetilde{q}(\vb) \widetilde{q}(\vc) }{ p_\tau( \vbeta, \sigma^2, \vb, \vc \mid \sD) } \right ]
$$
and
$$
\argmax_{\widetilde{q}} \text{ELBO}(\widetilde{q}) = \argmax_{\widetilde{q}} \bE_{\widetilde{q}} \left [ \log \left \{ \frac{p_\tau( \sD, \vbeta, \sigma^2, \vb, \vc)}{\widetilde{q}(\vbeta) \widetilde{q}(\sigma^2) \widetilde{q}(\vb) \widetilde{q}(\vc)} \right \} \right ].
$$
As pointed out in \cite{Zhang2020b}, the global KL-minimizer $q^{MF}$ is computationally intractable. Following the details in \cite{Ormerod2010}, the global minimizer may be approximated via coordinate ascent algorithm, where the approximation to $q_{\tau}^{MF}$ has the form
$$
q^{CAVI}(\vbeta) \propto \exp \left [ \bE_{-q(\vbeta)} \left \{ \log p_\tau( \sD, \vbeta, \sigma^2, \vb, \vc) \right \} \right ],
$$
where $\bE_{-q^{CAVI}(\vbeta)}$ denotes the expectation operator with respect to $q^{CAVI}(\sigma^2, \vb, \vc)$ while treating $\vbeta$ as fixed. Note that the CAVI approximate posterior at algorithmic convergence is only a local maximizer of the ELBO. The other CAVI density components have a similar form as above. In fact, we have:
\begin{align*}
	&q^{CAVI}(\vbeta) = N(\vmu_{\vbeta}, \mSigma_{\vbeta} ), \;\;\; q^{CAVI}(\sigma^2) = \text{InverseGamma} ( r_{\sigma^2}, s_{\sigma^2}), \\
	&q^{CAVI}(b_g) = \text{Gamma} ( r_{b_g}, s_{b_g}), \;\;\; q^{CAVI}(c_g) = \text{Exp} (s_{c_g})
\end{align*}
where we use the shape-rate gamma and inverse gamma distribution parameterization. The \textit{variational parameters} $\{ \vmu_\vbeta, \mSigma_\vbeta, r_{\sigma^2}, s_{\sigma^2}, r_{b_{1:G}}, s_{b_{1:G}}, s_{c_{1:G}} \}$ are computed using a coordinate ascent algorithm with details in Algorithm 1 of supplementary section F, and ``:" denotes contiguity of indices. The evidence lower bound (ELBO) is given by
\begin{align*}
	\text{ELBO}(\tau) &= \text{const} - \tfrac{p}{2} \log (\tau) - m_{1/\sigma^2} \left [ \frac{1}{2 \tau} \sum_{g=1}^G m_{b_g} \bE_q \lVert \vbeta_g \rVert^2 + \tfrac{1}{2} \bE_q \lVert \vy - \mX \vbeta \rVert^2 + s  \right ] + \tfrac{1}{2} \log \lvert \mSigma_\vbeta \rvert\\
	&- (r + \tfrac{n+p}{2}) \log(s_{\sigma^2}) - \sum_{g=1}^G \left \{ \log (s_{c_g}) + \tfrac{1}{2} (p_g + 1) \log(s_{b_g}) + \frac{m_{b_g} + 1}{s_{c_g}}   \right \},
\end{align*}
where
\begin{align*}
	&\bE_q \lVert \vbeta_g \rVert^2 = \lVert \vmu_{\vbeta; g} \rVert^2 + \tr(\mSigma_{\vbeta_g}) \\
	&\bE_q \lVert \vy - \mX \vbeta \rVert^2 = \lVert \vy - \mX \vmu_\vbeta \rVert^2 + \text{tr} \left ( \mX^\top \mX \mSigma_\vbeta \right ) \\
	&s_{b_g} = m_{c_g} + m_{1/\sigma^2} \frac{\lVert \vmu_{\vbeta_g} \rVert^2 + \text{tr}(\mSigma_{\vbeta_g}) }{2\tau} \\
	&m_{b_g} = \frac{(p_g+1)/2}{s_{b_g}} \\
	&s_{c_g} = 1 + m_{b_g} \\
	&m_{c_g} = 1/s_{c_g}.
\end{align*}
\noindent Details of the derivation of the CAVI component densities and the ELBO is provided in supplementary section F.

\subsection{Variational grouped penalized credible region}
Following the general group penalized credible region formulation in (\ref{ConstraintForm}), we replace the credible region 
based on the exact posterior $S_{\alpha,\pi}$ with the highest density region based on $t$ cycles of the CAVI algorithm:
\begin{equation}
	\label{SalphaThreshold}
	S_{\alpha}^{CAVI \, (t)} = \{ \vbeta \, : \, (\vbeta - \vmu_\vbeta^{(t)})^\top \mSigma_{\vbeta}^{^{(t)} \, -1} (\vbeta - \vmu_\vbeta^{(t)}) \le C^{(t)} \},
\end{equation}
where $\vmu_\vbeta^{(t)}$ and $\mSigma_{\vbeta}^{(t)}$ denote the value of CAVI approximate mean and variance of $\vbeta$ after $t$ cycles and $C^{(t)}$ is chosen such that the variational posterior probability of $S_{\alpha}^{CAVI \, (t)}$ equals to $1 - \alpha$. Hence, following the grouped horseshoe prior and CAVI posterior moments, our proposed estimator for $\vbeta$ can be expressed as
\begin{equation}
	\label{ConstraintFormMF}
	\widetilde{\vbeta}_{CAVI}^{(t)} =\argmin_{\vbeta \in \bR^{p}}  \sum_{g=1}^G \sqrt{p_g} \frac{u_g}{ \widehat{u}_g^2  },  \;\; \text{subject to} \;\; \vbeta \in S_{\alpha}^{CAVI \, (t)}.
\end{equation}
Since $S_{\alpha}^{CAVI \, (t)}$ is an elliptical credible region, we may express $\widetilde{\vbeta}_{CAVI}^{(t)}$ as the minimizer of a Lagrangian function
\begin{equation}
	\label{LagrangeForm}
	\widetilde{\vbeta}_{CAVI}^{(t)} =\argmin_{\vbeta \in \bR^{p}} (\vbeta - \vmu_\vbeta^{(t)})^\top \mSigma_{\vbeta}^{^{(t)} \, -1} (\vbeta - \vmu_\vbeta^{(t)}) + \lambda_\alpha \sum_{g=1}^G \sqrt{p_g} \frac{u_g}{ \widehat{u}_g^2  }.
\end{equation}
We refer to $\widetilde{\vbeta}_{CAVI}^{(t)}$ as our proposed variational grouped penalized credible region estimator (\texttt{VGPenCR}). The solution  $\widetilde{\vbeta}_{CAVI}^{(t)}$ is obtained using the procedure described in section \ref{subsec::Computations}, where $\widehat{\vbeta} = \vmu_\vbeta^{(t)}$ is the variational posterior mean and $\mSigma_{\vbeta}^{^{(t)}}$ is the variational posterior covariance after $t$ cycles. 
The corresponding set of selected groups is
$$
\sA_{CAVI}^{(t)} = \{ g \in [G] \, : \, \widetilde{u}_{g;CAVI}^{(t)} \neq 0 \},
$$
where $\widetilde{u}_{g;CAVI}^{(t)} = \lVert \widetilde{\vbeta}_{g;CAVI}^{(t)} \rVert$.
\\
\noindent Here, several key questions may be asked: does the use of the variational approximate posterior summaries lead to parameter and variable selection consistencies? And how many CAVI cycles $t$ do we need to ensure these consistency properties hold? Are more CAVI cycles necessarily better?

\section{Theoretical justification}
\label{sec::theory}
In this section we address the aforementioned questions by investigating the theoretical properties of our proposed post-processed Bayes estimator in (\ref{LagrangeForm}). We introduce the following notations for this article and supplementary sections B to E. Denote $[N] = \{1, \ldots, N\}$. For a stochastic sequence $\{ A_n \}$ and a non-negative deterministic sequence $\{ B_n \}$, we write $A_n = O_p( B_n )$ if for any $\gamma > 0$, there exists finite positive constants $N_\gamma$ and $\zeta_\gamma$ such that $ \bP \{ \lvert A_n \rvert > \zeta_\gamma B_n \} < \gamma $ for all $n \ge N_\gamma$.  Write $A_n = \Omega_p( B_n )$ if $B_n > 0$ and there exists $\zeta > 0$, $\gamma > 0$, and $N > 0$ such that  $ \bP \{ \lvert A_n \rvert \ge \zeta B_n \} \ge \gamma$ for all $n \ge N$. Write $A_n \asymp B_n$ if $A_n = \Omega_p(B_n)$ and $A_n = O_p(B_n)$. Write $A_n = o_p( B_n )$ if $\lim_{n \rightarrow \infty} \bP \{ \lvert A_n \rvert \ge \gamma B_n \} = 0$ for any $\gamma > 0$. We drop the subscript ``p" from our notations if $A_n$ is also a deterministic sequence. We say that a (random) sequence of estimators $\{ \widetilde{\vbeta}_n \}$ is parameter-consistent for a deterministic sequence $\{ \vbeta_n^0 \}$ if $\lVert \vbeta_n - \vbeta_n^0 \rVert = o_p(1)$. Define the corresponding selected groups as $\sA_n = \{  g \, : \, \widetilde{\vbeta}_{g;n} \neq \vzero \}$ and the true set as $\sA_n^0 = \{ g \, : \, \vbeta_{g;n}^0 \neq \vzero \}$,
where $\vbeta_{g;n}$ and $\vbeta_{g;n}^0$ are the $g$-th group of coefficients within $\vbeta_n$ and $\vbeta_n^0$ respectively. Say that $\{ \widetilde{\vbeta}_n \}$ is variable selection consistent for $\{ \vbeta_n^0 \}$ if $\lim_{n \rightarrow \infty} \bP^\star \{ \sA_n = \sA_n^0 \} = 1$.

Related theoretical results in the existing literature include those in \cite{Bondell2012}, \cite{Armagan2013}, and \cite{Zhang2018}. In particular, \cite{Bondell2012} showed that the non-grouped penalized credible region point estimator based on a normal prior is both parameter consistent and variable selection consistent. \cite{Armagan2013} showed that the exact non-grouped horseshoe posterior distribution for $\vbeta$ is consistent for $\vbeta^0$. \cite{Zhang2018} 
showed that the non-grouped penalized credible region point estimator based on a normal prior and Dirichlet-Laplace prior  consistent are both parameter consistent and variable selection consistent in a high-dimensional paradigm. However, it is worth noting that their results do not extend to grouped variable selection and pertain to exact posterior moments which can be computationally costly to estimate.

Here, we establish both the parameter consistency and variable selection consistency of our proposed estimator in (\ref{LagrangeForm}) for grouped-variable selection in the high-dimensional $p_n = o(n)$ paradigm based on the CAVI posterior.  To emphasize the dependence of various quantities with diverging $n$ in our notations, we introduce subscript $n$ whenever necessary. For example, the $n$ by $p_n$ design matrix of covariates is denoted as $\mX_n$. We denote the $j$-th largest eigenvalue of $\mX_n^\top \mX_n/n$ as $d_{j;n}$. We require the following assumptions:
\begin{enumerate}
	\item[(B1)] \textit{Divergence of $p$ and bounded group sizes}: For sample size $n$, the total number of predictors in the fitted model is $p_n = \sum_{g=1}^{G_n} p_g = o(n)$. Moreover, $\limsup_{n \rightarrow \infty} \max_{g \in [G_n] } p_g < \infty$.
	\item[(B2)] \textit{True model}: The true data generating distribution satisfies $y_i = \sum_{g \in \sA_n^0} \vx_{ig}^\top \vbeta_{g;n}^0 + \epsilon_i^\star$, where $\sA_n^0 \subseteq [G_n]$, and $\vepsilon^\star \sim N(\vzero, \sigma^{\star \; 2} \mPsi_n )$ with $\sigma^{\star \; 2} > 0$ and $\mPsi_n  = \mI_n - (1/n) \mJ_n$. The number of truly important predictors is $q_n = o_p (p_n)$. Moreover, the predictors are independent of the true random errors.
	\item[(B3)] \textit{Bounded and finite eigenvalues}: The sequence of smallest and largest eigenvalues of $\mX_n^\top \mX_n/n$ are bounded in the limit, i.e., $0 < d_{min} \le \liminf_{n \rightarrow \infty} d_{p_n;n } \le \limsup_{n \rightarrow \infty} d_{1;n} \le d_{max} < \infty$.
	\item[(B4)] \textit{Finite largest signal strength}: $\limsup_{n \rightarrow \infty} \max_{g = 1, \ldots G_n}  u_g^0  < \infty$, where $u_g^0 = \lVert \vbeta_{g;n}^0 \rVert$.
	\item[(B5)] \textit{Minimum true signal strength}: There exists two deterministic and diverging sequences $a_n^\star$ and $b_n$ such that $\sum_{g \in \sA_n^0} 1/u_g^{0} \le E_0 \sqrt{n / p_n}$ and $\sum_{g \in \sA_n^0} 1/(u_g^0)^{2} \le E_1 n /(p_n \sqrt{a_n^\star})$, for some $0 < E_0 < \infty$ and $0 < E_1 < \infty$, $C_n^{(t)} / (p_n a_n^{(t)}) \rightarrow c \in (0, \infty)$, $a_n^\star b_n p_n  = o(n)$ and $a_n^{(t)} = b_n^{1/(t+1)} a_n^\star$.
\end{enumerate}
(B1) requires the divergence in number of predictors to be sublinear. This divergence rate is assumed for establishing posterior consistency for many non-grouped global-local shrinkage priors \citep{Armagan2013} and is also assumed in \cite{Zhang2018}. (B2) is a sparsity assumption requiring the size of the number of significant predictors to grow at a slower rate than the total number of predictors. Note that (B2) differs from \cite{Armagan2013} and \cite{Zhang2018} where a more specific regime $q_n = o(n/\log(n))$ was assumed. (B3) requires $\mX_n^\top \mX_n/n $ to converge to a finite positive definite matrix. (B4) ensures that the largest true coefficient norm is bounded and thus implies that $\lVert \vbeta^0_n \rVert^2$ diverges at rate $q_n$. (B5) ensures that true signals are detectable, i.e., minimum norm among the significant groups is bounded below by $\sqrt{p_n/n}$, and prescribes a suitable divergence rate for the credible region threshold $C_n^{(t)}$, where we recall that $C_n^{(t)}$ is the threshold defined in (\ref{SalphaThreshold}). The $t$-dependent $C_n^{(t)}$ leads to a slow decreasing credible region size as the number of CAVI cycles increases. Note that following the prescription in (B5), the threshold $C_n^{(t)}$ decreases as number of cycles $t$ increases. This prescription is aligned with heuristics that as we increase the number of CAVI cycles, we should be more certain that the $\vmu_\vbeta^{(t)}$ is closer to $\vbeta_n^0$.

\noindent Our required results in this section hold under the following initialization of the CAVI algorithm, which we denote by ($\#$):
\begin{itemize}
	\item $m_{b_g}^{(0)}$: $0 \le \liminf_{g} m_{b_g}^{(0)} \le \limsup_{g} m_{b_g}^{(0)} < B < \infty$ and $1/\tau = O_p(\sqrt{a_n^{(0)}})$.
	\item $\vmu_\vbeta^{(0)}$: $\lVert  \vmu_\vbeta^{(0)} - \widehat{\vbeta}_{LS} \rVert^2 = o_p( n/ (a_n^{(0)} p_n))$, where $\widehat{\vbeta}_{LS} = (\mX_n^\top \mX_n)^{-1} \mX_n^\top \vy$.
	\item $\mSigma_\vbeta^{(0)}$: $R_{\min; n} \le \min\text{eigen} \{ \mSigma_\vbeta^{(0) \; -1} \} \le \max\text{eigen} \{ \mSigma_\vbeta^{(0) \; -1} \} \le R_{\max; n}$, where $R_{\min; n}$ and $R_{\max; n}$ are deterministic sequences such that $R_{\min; n} = \Omega ( a_n^\star b_n p_n )$.
	\item $m_{1/\sigma^2}^{(0)} =  (n - p) / \lVert \vy - \mX_n \widehat{\vbeta}_{LS} \rVert^2 + \xi_n$,  where $\xi_n = O_p(1)$.
\end{itemize}
Our first result establishes the parameter consistency and variable-selection consistency of our CAVI estimator.
\begin{theorem}
	\label{MFVBParameterVariableConsistency}
	Assume (B1) to (B5) holds. Under the initialization (\#), the estimator $\widetilde{\vbeta}_{CAVI}^{(t)}$ is parameter-consistent for $\vbeta^0$ for all $t \ge 2$. In particular,
	the order of stochastic convergence is:
	$$
	\lVert \widetilde{\vbeta}_{CAVI}^{(t)} - \vbeta^0_n \rVert \asymp  \sqrt{ \frac{a_n^{(t)} p_n}{n} } . 
	$$
	Moreover, $\widetilde{\vbeta}_{CAVI}^{(t)}$ is variable selection consistent for $\vbeta^0$ for all $t \ge 2$, i.e., $\bP \{ \sA_{CAVI}^{(t)} = \sA^0 \} \rightarrow 1$.
	\begin{proof}
		We provide an outline of our proof. First in Lemma 7 of supplementary section E, we show that $1/m_{1/\sigma^2}^{(t)}$ is a consistent estimator for $\sigma^2$, $\max_g m_{b_g}^{(t)}/\tau = O_p (\sqrt{a_n^{(0)}} ) = o_p(\sqrt{n})$. Next, we point out that $\vmu_{\vbeta}^{(t)}$ and $\mSigma^{(t)}$ satisfy the same conditions as generic posterior expectation and posterior covariance as defined in supplementary section B. Hence, $\widetilde{\vbeta}_{CAVI}^{(t)}$ may be considered a generic penalized credible region estimator.
		Finally, we quote Lemma 2 in supplementary section B and Lemma 6 in supplementary section C which states that the generic grouped penalized credible region estimator is both parameter-consistent and variable selection-consistent. Refer to supplementary sections B to E for details.
	\end{proof}
\end{theorem}
\noindent From Theorem \ref{MFVBParameterVariableConsistency}, we recommend running at least $2$ cycles to ensure that the resulting estimate is both parameter-consistent and variable selection-consistent. The next result describes the asymptotic improvement in estimator accuracy between successive CAVI cycles. Note that the exact stochastic convergence rate depends on the deterministic sequence $a_n^\star$ which bounds the convergence rate of the minimum true signal strength.
\begin{theorem}
	\label{LinearDecreasingNorm}
	Assume (B1) to (B5) holds. Under the initialization (\#), for any $t \ge 2$, we have
	$$
	\frac{ \lVert \widetilde{\vbeta}_{CAVI}^{(t+1)} - \vbeta^0_n \rVert }{ \lVert \widetilde{\vbeta}_{CAVI}^{(t)} - \vbeta^0_n \rVert } \xrightarrow{\bP^\star} 0.
	$$
	\begin{proof}
		This result follows from Theorem \ref{MFVBParameterVariableConsistency} and the fact that $a_n^{(t+1)}/a_n \xrightarrow{\bP^\star} 0$. Refer to supplementary section E for details. 
	\end{proof}
\end{theorem}
\noindent The preceding theorem implies that, provided sufficient computational resources are available to run at least two cycles, running additional CAVI iterations yield asymptotic improvement in estimation accuracy. This provides further justification to run at least two CAVI cycles.

\section{Numerical study}
\label{sec::numericalStudy}

We conduct simulation studies for \texttt{VGPenCR} to assess its variable selection and estimation performance in three simulation examples corresponding to different models: (1) Generalized additive model; (2) Regression with categorical variables; (3) Nonparametric varying-coefficient model. In each study, we compare the performance of \texttt{VGPenCR} with four other approaches:
\begin{itemize}
	\item \texttt{VPenCR}: Variational penalized credible region  with grouped horseshoe prior and non-grouped optimization objective.
	\item \texttt{PenCRDL}: Penalized credible region with Dirichlet-Laplace global-local shrinkage prior \citep{Zhang2018}
	\item \texttt{SSGL}: Spike-and-slab group Lasso \citep{Bai2022}
	\item \texttt{BGLSS}: Bayesian group Lasso \citep{Xu2015}
\end{itemize}

In all examples, the CAVI algorithm is run until the change in ELBO is less than 0.01\%. The \texttt{VPenCR} estimate is obtained by applying the non-grouped penalized credible region method \citep{Bondell2012} to the variational posterior mean and covariance matrix. The \texttt{PenDRDL} estimate is obtained by applying non-grouped penalized credible region method to the posterior mean and covariance from running 10000 Gibbs sampler iterations using the R package \texttt{dlbayes}, discarding the first 5000 iterations as burn-in. For \texttt{VGPenCR}, \texttt{VPenCR} and  \texttt{PenCRDL}, the shrinkage parameter $\lambda_\alpha$ is chosen by cross validation. The method \texttt{SSGL} is implemented using the R package \texttt{SSGL} with convergence threshold 0.01. For the non-grouped penalized credible region methods, a group is selected if at least one of its elements is deemed important. The method \texttt{BGLSS} is implemented by the R package \texttt{MBSGS}, running 10000 Gibbs sampler iterations and keeping the last 5000 iterations. All \texttt{R} packages are available on \texttt{CRAN}.

\paragraph{Performance metrics}

In each simulation study, we compare the methods with the following performance metrics:

\begin{itemize}
	\item Youden's index 
	$$ J = \frac{\text{TP}}{(\text{TP} + \text{FN})} + \frac{\text{TN}}{(\text{TN} + \text{FP})}- 1	, $$
	where TN, TP, FN, and FP denote the number of true negatives, true positives, false negatives, and false positives respectively.
	\item Matthew's correlation coefficient
	$$\text{MCC} = \frac{(\text{TP}\times\text{TN})-(\text{FP}\times\text{FN})}{\sqrt{(\text{TP}+\text{FP})(\text{TP}+\text{FN})(\text{TN} + \text{FP})(\text{TN}+\text{FN})  }}. $$
	\item Given test points $\{ \breve{\vx_i}^\star \}_{i \in \sI_{test}}$, the mean squared prediction error is
	$$\text{MSPE} = \frac{1}{\lvert \sI_{test} \rvert } \sum_{i \in \sI_{test}} (\breve{y}_i^\star - \widetilde{\mu}_i )^2, $$	
	where $\widetilde{\mu}_i$ is the predicted response of the $i$-th (uncentered) test point from using the estimated coefficients by each of the competing approaches. 
	\item Running time in seconds.
\end{itemize}

The Youden's index $J$ and Matthew's correlation coefficient MCC represent the variable selection performance, whilst the mean squared prediction error MSPE is a measure of the out-of-sample prediction accuracy. The running time is an indicator of the computational cost of each approach. All examples were completed on a MacBook Pro M4 with 24GB memory. 

\subsection{Generalized additive model}
\label{sim::GAM}
We generate 100 datasets of size $n= 200$ observations from the data-generating process
$$\breve{y}_i  = 5\sin (\pi z_{i1}) + 2.5(z_{i3}^2 - 0.5) + e^{z_{i4}} + 3z_{i5}  + \breve{\epsilon}_i,\quad \breve{\epsilon}_i \sim N(0,1),$$
where $z_{ig} \sim U(0,1)$ and $G \in \{ 50, 100, 150 \}$.
\noindent The mean response $\bE(\breve{y}_i|\breve{\vx}_i)$ is modeled as 
$$\bE(\breve{y}_i|\vz_i)  = \breve{\vx}_{i1} \boldsymbol{\beta}_1 + \dots + \breve{\vx}_{iG} \boldsymbol{\beta}_G, $$
where $\breve{\vx}_{ig} :=  \breve{\vx}(z_{ig})$ denotes a natural cubic splines basis with dimension 4. For each dataset, we generate a test dataset containing $200$ test points to compute the MSPE.

\begin{figure}[h!]
	\centering
	\includegraphics[width=1.0\linewidth]{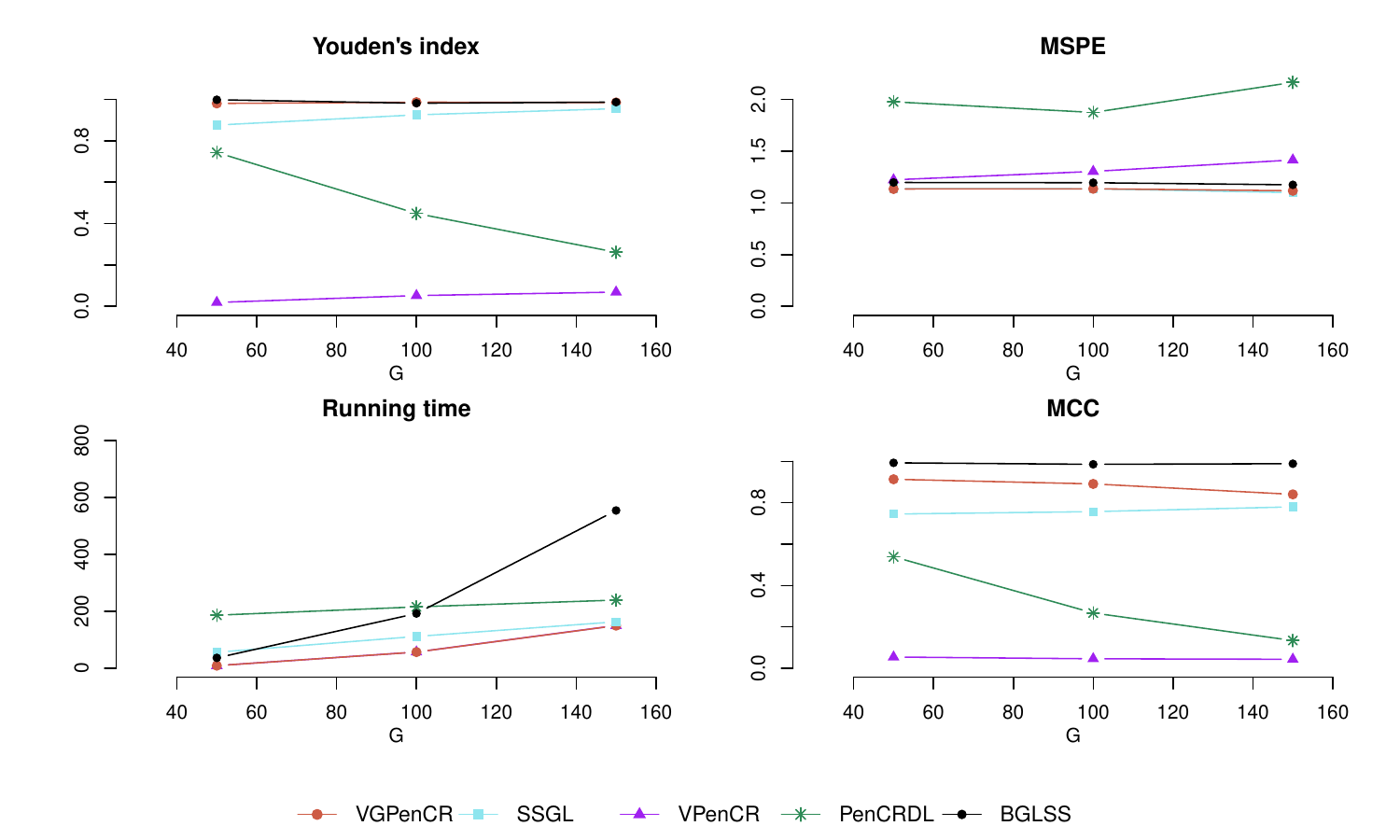}
	\caption{Youden's Index, MSPE, running time and Matthew's correlation coefficient of \texttt{VGPenCR}, \texttt{SSGL},  \texttt{VPenCR}, \texttt{PenCRDL} and \texttt{BGLSS} for  $G = 50, 100, 150$. The numbers are averages over 100 replications.}
	\label{fig:example1}
\end{figure}

The results are summarized in Figure \ref{fig:example1}. Overall, \texttt{VGPenCR} and \texttt{BGLSS} are the two superior methods in this example. Clearly, \texttt{VGPenCR} exhibits better performance over the two non-grouped penalized credible region methods (\texttt{VPenCR} and \texttt{PenCRDL}). The poorer variable selection performance by the non-grouped penalized credible region methods is attributed to an inflated false positive rate that is consequential to the lack of a group-sparsity effect. This performance disparity is evidence that grouped sparsity enforcement as per equation \eqref{ConstraintForm} is indeed necessary.

Notably, \texttt{VGPenCR} and \texttt{BGLSS} demonstrate comparable Youden's index and MSPE.  Moreover, the running time of \texttt{VGPenCR} is  3.4 to 4 times shorter than \texttt{BGLSS}. In fact, 
both \texttt{VPenCR} and \texttt{VGPenCR} require substantially lower computation time than the other competing methods.

\subsection{Regression with categorical variables}
\label{sim::CatPred}
In this example, we consider a linear regression model with $K$ categorical predictors, each with three levels $1$, $2$, and $3$. We generate 100 datasets with $K \in \{ 10, 12, 15 \}$ categorical predictors and $n = 200$. For each dataset, we generate a test dataset containing $200$ test points to compute the MSPE. The data-generating process is
\begin{align*}
	\breve{y}_i &= 2I(z_{i1}=2) - I(z_{i1}=3) + 4.5I(z_{i2}=2) + 5I(z_{i2}=3) + 1.5I(z_{i1}=2,z_{i2}= 2) \\ &- 3.5I(z_{i1}=2,z_{i2}= 3) + 2I(z_{i1}=3,z_{i2}= 2) + 4I(z_{i1}=3,z_{i2}= 3) + \breve{\epsilon}_i,\\ 
	z_{i1} &\sim \text{Categorical}( \{ 1, 2, 3\}, \vp = (0.3, 0.65,0.05)  ) \\
	z_{i3} &\sim \text{Categorical}( \{ 1, 2, 3\}, \vp = (0.2, 0.5,0.3)  ) \\
	z_{i4} &\sim \text{Categorical}( \{ 1, 2, 3\}, \vp = (0.5, 0.2,0.3)  ) \\
	z_{ij} &\sim \text{Categorical}( \{ 1, 2, 3\}, \vp = (0.33, 0.33,0.33)  ), \;\; j = 2, 5, \ldots, K, \\
	\breve{\epsilon}_i & \sim N(0,1).
\end{align*}

\noindent Note that only the main effects of $z_{i1}$ and $z_{i2}$ and their interaction are truly important. For each data set, we fit a regression with all $K$ main effects and $\binom{K}{2}$ two-way interaction terms. Note that each main effect is represented as a group of 2 covariates and each interaction term is represented as a group of 4 covariates. In total, we have $G = 55, 78$, and $120$ groups that correspond to the three different settings of $K$.

\begin{figure}[h!]
	\centering
	\includegraphics[width=1.0\linewidth]{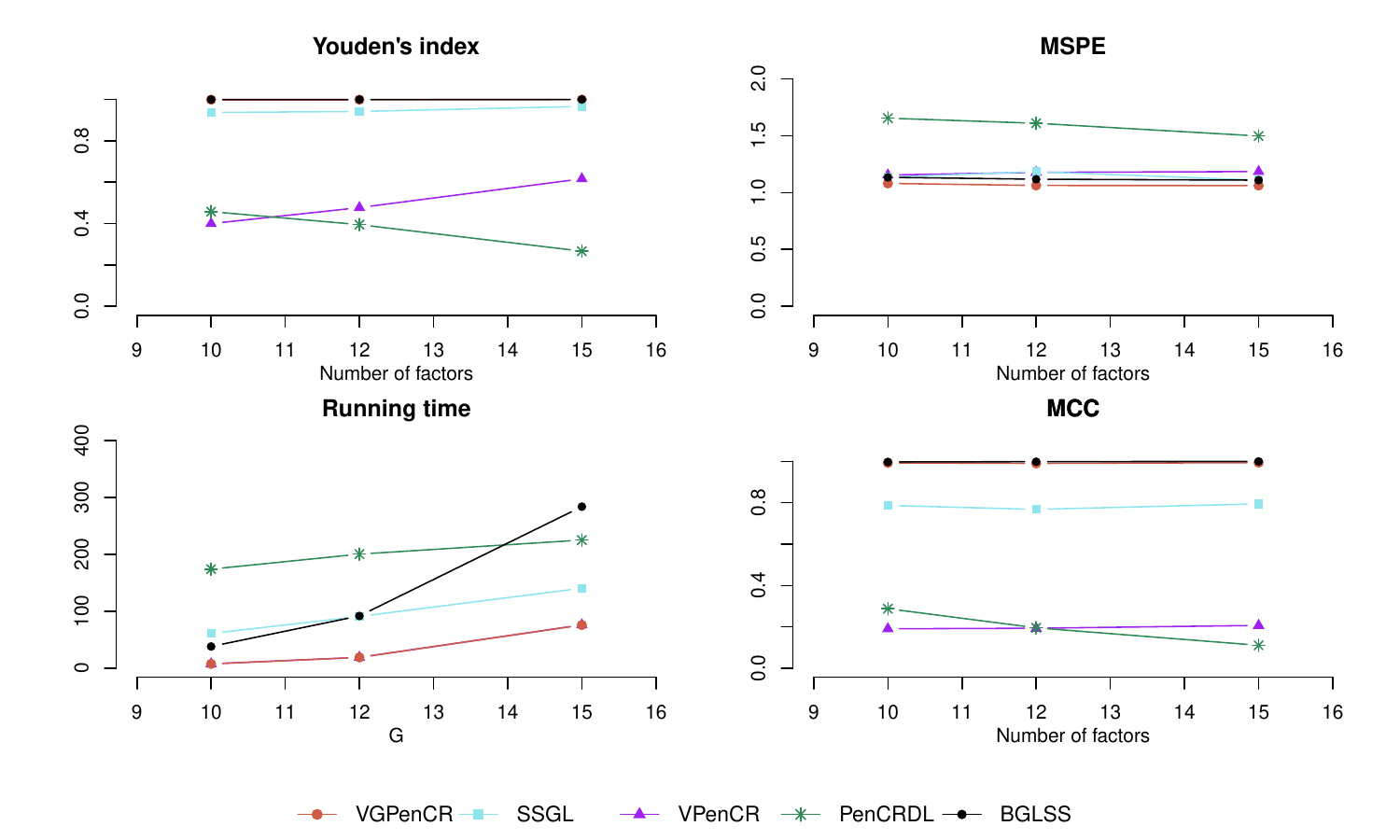}
	\caption{Youden's Index, MSPE, running time and Matthew's correlation coefficient of \texttt{VGPenCR}, \texttt{SSGL},  \texttt{VPenCR}, \texttt{PenCRDL} and \texttt{BGLSS} for  $K = 10, 12, 15$. The numbers are averages over 100 replications.}
	\label{fig:example2}
\end{figure}

The results are summarized in Figure \ref{fig:example2}. Overall, \texttt{VGPenCR} is the best method in this example. In terms of variable selection, \texttt{VGPenCR} and \texttt{BGLSS} have similar performance and are better than the other methods. Moreover, \texttt{VGPenCR} has the lowest MSPE and running time among all methods. In particular, similar to the results in \ref{sim::GAM}, \texttt{VGPenCR} exhibits better performance across all metrics compared to \texttt{VPenCR} and \texttt{PenCRDL}.

\subsection{Nonparametric varying-coefficient model}
\label{sim::NonparaVC}
Nonparametric varying-coefficient models are useful for modeling time-varying effects on responses that are measured repeatedly. The following example is adapted from \cite{Bai2023}. We generate 100 datasets with $n = 50$ and $G \in \{ 30, 50, 80 \}$  from the true data-generating process:
\begin{align*}
	&\breve{y}_i(t_{ij}) = \sum_{g =1}^G \breve{x}_{ig}(t_{ij})\beta_g^0(t_{ij})  + \breve{\epsilon}_{ij},\; i = 1, \dots, n, \; j = 1,\dots, m_i,\\
	&\breve{\epsilon}_{ij} \sim N(0,1),
\end{align*}
where
\begin{align*}
	& \beta_1^0(t)= 10\sin \left( \frac{\pi t}{15} \right), \, \beta_2^0(t) = -0.6t + 6, \, \beta_3^0(t) = -1 + 2\sin \left(\frac{\pi(t-25)}{8} \right), \\
	& \beta_4^0(t) = 1 + 2\cos \left(\frac{\pi(t-25)}{15}\right), \, \beta_5^0(t) = 2 + \frac{10}{1+e^{10-t}}, \, \beta_6^0(t) = -5, \\
	&\beta_7^0(t) = \cdots = \beta_G^0(t) = 0. 
\end{align*}
Note that $\beta_1^0(t), \beta_3^0(t), \beta_4^0(t)$, and $\beta_5^0(t)$ are non-linear functions. The observation time $t_{ij}$ is sampled from $\{1,2,\dots,20\}$ and each time point has a 60\% chance of being skipped. We also add a random perturbation from $U(-0.5,0.5)$ to each non-skipped time point, in order to obtain irregularly-spaced data. The first covariate $\breve{x}_{i1}$ is simulated from $U(t/10,2+t/10)$. Conditioned on $\breve{x}_{i1}$, the covariates $\breve{x}_{ik}$, $k = 2,\dots, 5$ are i.i.d drawns from $N(0,(1+\breve{x}_{i1})/(2+\breve{x}_{i1}))$. The covariate $\breve{x}_{i6}$ is drawn from $N(1.5e^{t/40},1)$. For each $i$, $\breve{\vx}_{i;7:G}$ is drawn from a multivariate normal distribution with mean $\vzero$ and covariance structure $\text{cov}(\breve{x}_{ik}(t),\breve{x}_{ik}(s)) = 0.5^{\vert t - s \vert}$. We fit the following nonparametric varying coefficient model with $G$ covariates
$$\breve{y}_i(t_{ij}) = \sum_{g =1}^G \breve{x}_{ig}(t_{ij})\beta_g(t_{ij})  + \breve{\epsilon}_{ij},\; i = 1, \dots, n, \; j = 1,\dots, m_i,$$
where each $\beta_g$ is modeled by a linear combination of a $d$-dimensional B-splines basis
\[ \beta_g(t) \approx \sum_{l=1}^d \gamma_{gl}B_{gl}(t),\]
and $B_{gl}(t)$ are B-splines basis functions with equispaced knots. Here, we set $d=8$. The fitted model may be reformulated in concise matrix form. Define 
$\mB_{g\cdot} (t) = (B_{g1} (t), \ldots, B_{gd} (t))^\top$ and the basis expansion matrix as
$$
\mB(t) = \begin{pmatrix}
	\mB_{1\cdot} (t) & \vzero & \ldots & \vzero \\
	\vzero & \mB_{2\cdot} (t) & \ldots & \vzero \\
	\vdots & \vdots & \vdots & \vdots \\
	\vzero & \vzero & \ldots & \mB_{G\cdot} (t) \\
\end{pmatrix} \in \bR^{Gd \times G}.
$$
Denote $\breve{\vx}_{i \cdot} ( t_{ij} ) = (\breve{x}_{i1} (t_{ij}) , \ldots,\breve{x}_{iG} (t_{ij})  )^\top$ and $\mU_i = (\vu_{i1}, \ldots, \vu_{i m_i})^\top$, where
\begin{align*}
	\vu_{ij} &= ( B_{11} (t_{ij}) \breve{x}_{i1} (t_{ij}), \ldots, B_{Gd} (t_{ij}) \breve{x}_{iG} (t_{ij}))^\top \\
	&=  \mB(t_{ij}) \breve{\vx}_{i \cdot} ( t_{ij} ) \in \bR^{Gd}
\end{align*}
Then, by denoting $\breve{\vepsilon}_i = ( \breve{\epsilon}_{i1}, \ldots, \breve{\epsilon}_{i t_{m_i}} )$, $\breve{\vy}_i  = (\breve{y}_i(t_{i1}), \ldots,  \breve{y}_i(t_{im_i}) )^\top$, $\vgamma = (\vgamma_{1\cdot}^\top, \ldots, \vgamma_{G\cdot}^\top)^\top$, and $\vgamma_{g \cdot} = ( \gamma_{g1}, \ldots, \gamma_{gd} )^\top$, we have
$$
\breve{\vy}_i =  \mU_i \vgamma + \breve{\vepsilon}_i.
$$
Stacking the $\breve{\vy}_i$'s and $\breve{\vepsilon}_i$'s into their respective $N = \sum_{i} m_i$ dimensional vector, we have
$$
\breve{\vy} = \mU \vgamma + \breve{\vepsilon},
$$
where $\breve{\vepsilon} \sim \text{MVN}(\vzero, \sigma^2 \mI_N)$ and $N = \sum_{i} m_i$. We enforce group-level sparsity for \texttt{VGPenCR} and \texttt{BGLSS} by shrinkage and sparsification of the coefficient groups $\{ \vgamma_{g \cdot}  \}$.  In particular, our proposed \texttt{VGPenCR} approach assigns the following prior on the coefficient groups:
$$
\vgamma_{g\cdot} \vert \sigma^2, \tau, b_g \sim \text{MVN}(\vzero, \tau \sigma^2/b_g \mI_{d}), \; b_g \sim \text{Ga}(1/2, c_g), \; c_g \sim \text{Ga}(1/2,1)
$$ 
and $\sigma^2 \sim \text{InvGa}(r,s)$. Note that this group-level sparsity enforcement reflects our assumption that most of the coefficient functions $\beta_j^0(\cdot)$ are constant at $0$. To assess the accuracy of each method for estimating the time-varying coefficient functions, we compute the Mean Integrated Squared Error (MISE) for each $\beta_g$:
$$\text{MISE}_g  = \int_0^{20} (\beta_g^0(t)- \widehat{\beta}_g(t))^2 dt$$
In practice, since the MISE involves an integral, we use numerical integration via trapezoid rule as an approximation.

\begin{figure}[h!]
	\centering
	\includegraphics[width=0.8\linewidth]{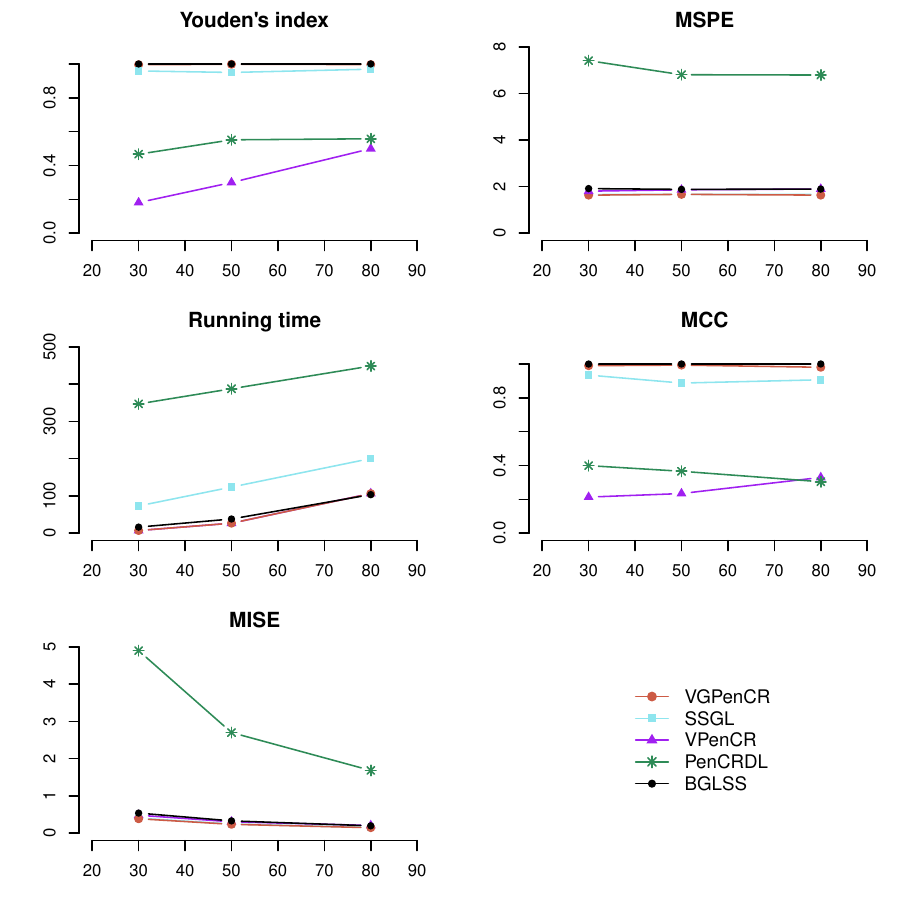}
	\caption{Youden's Index, MSPE, running time, Matthew's correlation coefficient and MISE of \texttt{VGPenCR}, \texttt{SSGL},  \texttt{VPenCR}, \texttt{PenCRDL} and \texttt{BGLSS} for  $G = 30, 50, 80$. The numbers are averages over 100 replications. }
	\label{fig:example3}
\end{figure}
\begin{figure}[h!]
	\centering
	\includegraphics[width=1.0\linewidth]{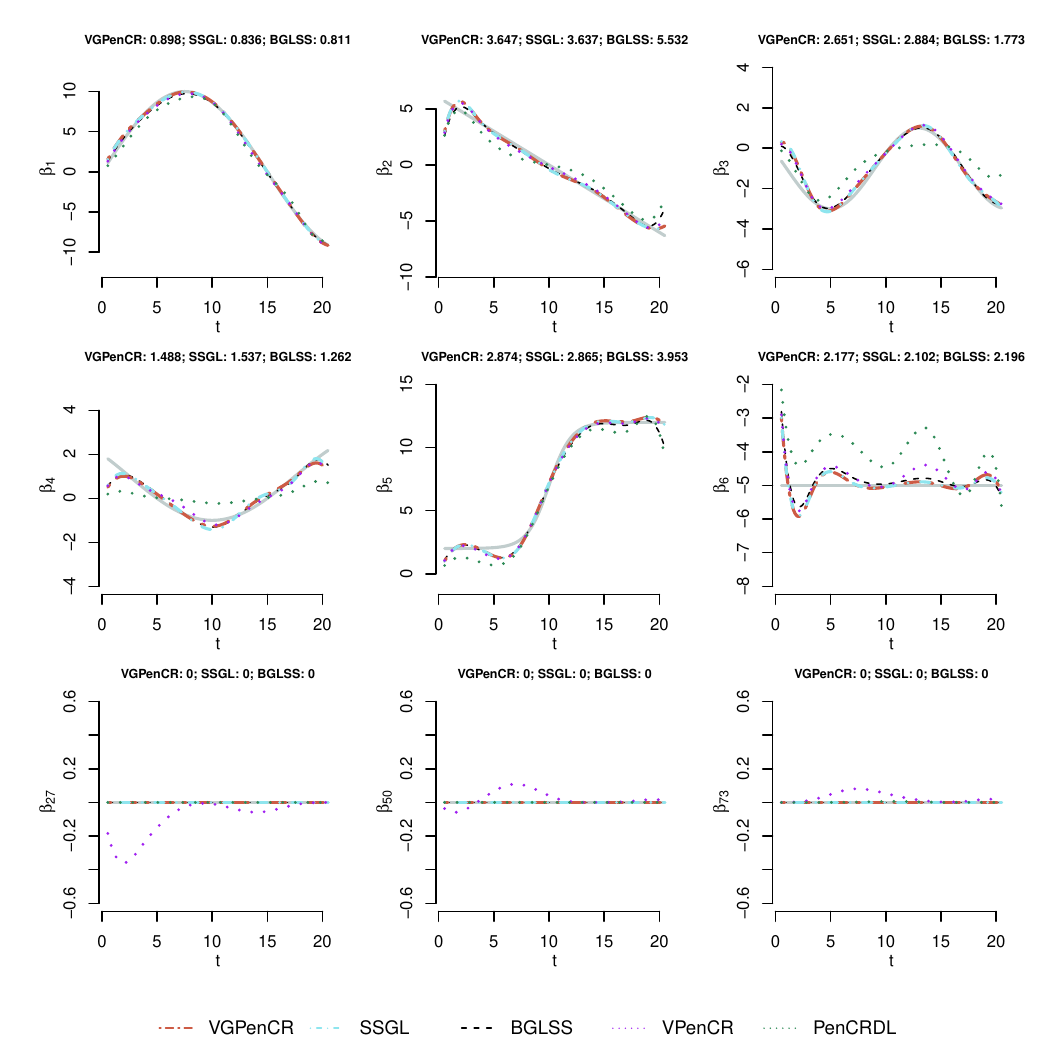}
	\caption{MISE of selected time varying coefficients $\beta_g(\cdot)$ for $g = \{1,\dots, 6, 27, 50, 73\}$ in one dataset with $G= 80$ . The numbers in their respective plot titles are MISE$_g$ of \texttt{VGPenCR}, \texttt{SSGL}, \texttt{BGLSS}. }
	\label{fig:example3Beta}
\end{figure}

The results are summarized in Figures \ref{fig:example3} and \ref{fig:example3Beta}. Overall, \texttt{VGPenCR} and \texttt{BGLSS} are the two superior methods in this example. More specifically, \texttt{VGPenCR} has the best MISE, MSPE, and running time, whereas \texttt{BGLSS} has the best Youden's index and MCC. Similar to previous examples, \texttt{VGPenCR} exhibits better performance across all metrics when compared to \texttt{SSGL}, \texttt{VPenCR} and \texttt{PenCRDL}. Figure \ref{fig:example3Beta} shows the estimated functions $\beta_g(\cdot)$ for selected coefficient functions from one simulation for $G = 80$. It can be seen from both figures that except for \texttt{PenCRDL} and \texttt{VPenCR}, all other methods can approximate the non-linear time-varying coefficients quite accurately, which suggests that enforcing grouped sparsity leads to better variable selection and prediction performance.

\newpage

\section{Application to NHANES data }
\label{sec::NHANESexample}

In this example, we examine the performance of our \texttt{VGPenCR} in identifying important main effects and interactions of environmental exposures. We analyze the data from the 2001-2002 cycle of the National Health and Nutrition Examination Survey (NHANES). The response variable of interest here is the leukocyte telomere length (LTI), which is commonly used as a proxy for overall telomere length. Previous studies suggested that organic pollutants can be associated with telomere length \citep{mitro2015cross}, and our goal is to identify the organic pollutants that are associated with changes in LTI levels. This data was previously analyzed by \cite{antonelli2020estimating} and \cite{Bai2022}.

We follow \cite{Bai2022} and consider 18 organic pollutants (also called \textit{exposures}), together with 13 demographic variables. Note that all organic pollutants measurements are continuous covariates. The log of LTI is modeled with a generalized additive model in Section \ref{sim::GAM} involving $G=184$ groups. In particular, the model is formulated as:
$$y_i = f_{\vbeta}(\vz_i) + \epsilon_i, \,\, \epsilon\sim N(0,\sigma^2)$$
where $y_i \in \bR$ is the scaled log LTI, $\vz_i$ is a $31$-dimensional vector of covariate information, and 
$$f_{\vbeta}(\vz_i) = \breve{\vx}_{i1}(\vz_{i}) \boldsymbol{\beta}_1 + \dots + \breve{\vx}_{iG}( \vz_{i}) \boldsymbol{\beta}_G. $$
Details of the predictor groups $\breve{\vx}_{ig}$ are as follows. The first eighteen predictor groups $\{ \vx_{ig} (\vz_i) \}_{g=1}^{18}$ are two-dimensional vectors that represent the splines bases for the 18 organic pollutants. The next 153 groups $\{ \vx_{ig} (\vz_i) \}_{g=19}^{171}$ are four-dimensional vectors that represent the interaction between each pair of the 18 groups of splines bases corresponding to the organic pollutants. The remaining 13 groups $\{ \vx_{ig} (\vz_i) \}_{g=172}^{184}$ have dimensions between one to three, where they represent each of the 13 demographic variables. The final dataset has $n=1003$ observations with $p=18\times2 + 4\times {18\choose 2} + 18  = 666$ coefficients.

Similar to the simulation study, the CAVI algorithm for \texttt{VGPenCR} is run until the change in ELBO is less than $0.01\%$. \texttt{BGLSS} is implemented by the R package \texttt{MBSGS} and \texttt{SSGL} is implemented using the R package \texttt{SSGL}. The \texttt{PenCRDL} estimate is obtained by applying the non-grouped penalized credible  region method to the posterior mean and covariance 3000 post burn-in Gibbs sampler iterations using the R package \texttt{dlbayes}. For all methods except \texttt{BGLSS}, the shrinkage parameters are chosen by cross validation. 

We examine the performance of \texttt{VGPenCR} with \texttt{SSGL}, \texttt{BGLSS} and \texttt{PenCRDL} by 10-fold cross validation. We repeat the cross-validation on 100 different random partitions of the data. This experiment allows us to thoroughly assess  all methods in terms of their ability to identify the effects, and compare their performances in terms of predictive accuracy and computational time. Table \ref{tab:CVTimeNHANES} summarizes the result of 10-fold cross validation on 100 different partitions of the data. Among all methods, \texttt{VGPenCR} has the lowest median cross validation error. A sign test confirms that \texttt{VGPenCR} has significantly lower cross-validation error than all other competing methods (p-value = 0.000895). In addition, \texttt{VGPenCR} has the shortest running time on average.

In terms of variable selection, all methods select age as an important predictor throughout all cross-validation iterations. Since \texttt{PenCRDL} does not impose group variable selection, it tends to select much more predictors than the other approaches. On the other hand, \texttt{BGLSS} mostly only selects age. In addition to age, gender, and a pollutant (Furan1) are often selected by \texttt{SSGL}, but the method did not pick up any interaction term. \texttt{VGPenCR} often identified PCB169 and some interactions as important. The summary of the selection rate of the 18 pollutants and their interactions and the demographic variables over the 100 cross-validation iterations can be found in the Supplementary Material. \footnote{Unlike \cite{Bai2022}, we also perform variable selection on the 18 demographic variables in this example.}

\begin{table}
	\centering
	\begin{tabular}{ccccc}
		\hline
		& VGPenCR & SSGL & BGLSS  & PenCRDL \\ \hline
		Median & 0.7756 & 0.7785  & 0.7807 & 0.7911\\ 
		IQR   & 0.0039 & 0.0017  & 0.0036 & 0.0192 \\ 
		Running time  & 209.95 & 397.41  & 999.28 & 6605.51\\ \hline
	\end{tabular}
	\caption{Median and IQR of the cross-validation error over the 100 repetitions for each method. The last row is the average computational time for one cross validation fold of all methods.}
	\label{tab:CVTimeNHANES}
\end{table}


\section{Conclusion}
\label{sec::ConclusionFutureWork}

We propose a fast and accurate approach to grouped variable selection for several high-dimensional Bayesian regression models by post-processing posterior summaries. Our proposed method accurately identifies significant groups through a grouped penalized credible region framework that post-processes posterior summaries. To achieve low computational cost, our method avoids estimating posterior summaries via potentially time-consuming stochastic sampling, and instead approximates these summaries using CAVI. The resultant estimator exhibits good theoretical properties including parameter consistency and variable selection consistency. Moreover, our theoretical analyses provides practical recommendations on the minimum number of CAVI cycles to run. Through a numerical study, our proposed method exhibits strong variable selection and predictive performance in both simulated and real datasets, often outperforming competing state-of-the-art approaches to Bayesian high-dimensional regression. Moreover, the numerical results confirm the importance of jointly incorporating a grouped shrinkage prior and grouped sparsity penalization within the proposed framework. A limitation of our work is that the proposed framework does not provide uncertainty quantification for the coefficient vector. Nevertheless, it already meets the modeling needs in a wide range of scenarios where the primary goals are grouped variable selection and accurate prediction.

In summary, the proposed \texttt{VGPenCR} is a computationally efficient and accurate method for grouped variable selection and prediction in Bayesian high-dimensional regression models.

$\text{   }$


\section*{Acknowledgements} 
The authors do not have any competing interests to declare. The first author was partially funded by the \textit{Australian Research Council}.



\bibliographystyle{elsarticle-harv}
{

	\setlength{\itemsep}{0em}
	\setlength{\parskip}{0em}
	\bibliography{GroupPenCRbibli}
}

\newpage



\renewcommand{\thesection}{\Alph{section}}
\renewcommand{\thesubsection}{\thesection.\arabic{subsection}}

\setcounter{section}{0}
\setcounter{equation}{0}
\setcounter{table}{0}
\setcounter{section}{0}
\setcounter{figure}{0}
\setcounter{page}{1}
\setcounter{footnote}{0}
\setcounter{lemma}{0}
\newcolumntype{L}[1]{>{\raggedright\arraybackslash}p{#1}}
\newcolumntype{C}[1]{>{\centering\arraybackslash}p{#1}}
\newcolumntype{R}[1]{>{\raggedleft\arraybackslash}p{#1}}

\def\spacingset#1{\renewcommand{\baselinestretch}%
	{#1}\small\normalsize} \spacingset{1}

\newgeometry{left=0.5in, right = 0.5in,bottom=0.5in, top = 0.5in}

\if0\blind
{
	\bigskip
	\bigskip
	\bigskip
	\begin{center}
		{\LARGE\bf Supplementary material for: \\ Variational approximate penalized credible regions for Bayesian grouped regression}
	\end{center}
	\medskip
} \fi

\bigskip

\noindent%


\spacingset{1.5}

    \section{Justification for using a group lasso algorithm to compute proposed minimizer}
To see why $\widetilde{\vbeta}$ may be computed using a group lasso algorithm (such as the R package \texttt{gglasso}), note that the package includes codes to compute the penalized least squares minimizer
$$\vbeta^* = \argmin_{\vbeta \in \bR^{p}} (\mY^* - \boldsymbol{X}^*\vbeta)^\top (\mY^* - \boldsymbol{X}^*\vbeta) + \lambda_\alpha \sum_{g=1}^G \sqrt{p_g} \lVert \vbeta_g \rVert_2.$$
Observe that
\begin{align*}
	\vbeta^* & = \argmin_\vbeta (\mY^* - \boldsymbol{X}^*\vbeta)^\top (\mY^* - \boldsymbol{X}^*\vbeta) + \lambda_\alpha \sum_{g=1}^G \sqrt{p_g} \lVert \vbeta_g \rVert_2 \\
	& =\argmin_\vbeta (\mSigma_\vbeta^{-1/2}\widehat{\vbeta} - \mSigma_\vbeta^{-1/2}\boldsymbol{D}\vbeta)^\top(\mSigma_\vbeta^{-1/2}\widehat{\vbeta} - \mSigma_\vbeta^{-1/2}\boldsymbol{D}\vbeta) + \lambda_\alpha \sum_{g=1}^G \sqrt{p_g} \lVert \vbeta_g \rVert_2 \\
	& = \argmin (\widehat{\vbeta} - \boldsymbol{D}\vbeta)^\top \mSigma_\vbeta^{-1}(\widehat{\vbeta} - \boldsymbol{D}\vbeta) + \lambda_\alpha \sum_{g=1}^G \sqrt{p_g} \lVert \vbeta_g \rVert_2 \\
	& = \argmin_\vbeta (\boldsymbol{D}\vbeta - \widehat{\vbeta})^\top \mSigma_\vbeta^{-1} (\boldsymbol{D}\vbeta - \widehat{\vbeta}) + \lambda_\alpha \sum_{g=1}^G \sqrt{p_g} \lVert \vbeta_g \rVert_2  \\
	&= \mD^{-1} \widetilde{\vbeta}.
\end{align*}

\section{Primer to proofs}
In primer sections B and C, we establish the parameter and variable selection consistency for a class of \textit{generic grouped penalized credible region} (\texttt{GroupPenCR}) \textit{estimators}. Here, the generic \texttt{GroupPenCR} estimator is defined as 
\begin{equation}
	\label{genericGroupPenCREstimatorForm}
	\widetilde{\vbeta}_n = \argmin_{\vbeta \in \sS} \sum_{g=1}^G \frac{u_g}{  \widehat{u}_g^2 }
\end{equation}
based on an elliptical \textit{generic credible region} of the form:
\begin{equation}
	\label{genericCredibleRegion}
	\sS = \{ \vbeta \, : \, (\vbeta - \widehat{\vbeta}_n)^\top \widehat{\mSigma}^{-1} (\vbeta - \widehat{\vbeta}_n) \le C_n \},
\end{equation}
where $C_n \asymp a_n p_n$ for some possibly diverging sequence $a_n = o( n / p_n)$, the \textit{generic posterior expectation} is
\begin{equation}
	\label{eqn::genericBetaHat}
	\widehat{\vbeta}_n = \left \{  \mX_n^\top \mX_n + \mDelta_n \right \}^{-1} \mX_n^\top \vy
\end{equation}
and the \textit{generic posterior covariance matrix} is
\begin{equation}
	\label{eqn::genericSigma}
	\mSigma = \widetilde{s}_n \left ( \mX_n^\top \mX_n + \mDelta_n \right )^{-1},
\end{equation}
for some consistent estimator $\widetilde{s}_n \xrightarrow{\bP^\star} \sigma^2$ and $\mDelta_n = \text{BlockDiag} \left \{ \delta_{1} (\vy) \mI_{p_1}, \ldots, \delta_{G_n} (\vy) \mI_{p_G} \right \}$, and each $\{\delta_{g} (\vy) \}_{g \ge 1}$ is a non-negative stochastic sequence that satisfies
$$
\frac{\max_{g=1,\ldots,G_n} \delta_{g} (\vy)^2 }{n } \xrightarrow{\bP^\star} 0
$$
or equivalently $\max_{g \ge 1} \delta_{g} (\vy) = o_p(n^{1/2})$. Here, we may interpret $\widehat{\vbeta}$ and $\mSigma$ as the posterior mean and posterior covariance of $\vbeta$ based on the data-dependent prior:
\begin{equation}
	\label{eqn::highDimNormalPrior}
	\vbeta \mid \mDelta_n \; \sim \; \text{N}(\vzero, \sigma^2 \mDelta_n).
\end{equation}
Note that throughout our proofs, we write $\sigma^2$ for $\sigma^{\star \; 2}$ when there is no ambiguity.
\section{Parameter consistency of generic GroupPenCR estimator}
We prove that $\vbeta_n^0$ is contained in the generic credible region (\ref{genericCredibleRegion}) with probability approaching 1. While the proofs are extensions of those presented in \cite{Zhang2018} to the grouped variable selection setting, one important novel contribution here is that we establish a stronger statement regarding asymptotic behavior $\lVert \widetilde{\vbeta}_n - \vbeta^0_n \rVert$ in terms of ``$\asymp$", whereas \cite{Zhang2018} presented their results in term of ``$O_p$". The stronger statement we proved here is necessary for us to prove Theorem 2 in section E.
\begin{lemma}
	\label{ContainsTruthProbApproachOne}
	Suppose (B1) to (B4) hold and that $C_n/p_n \rightarrow \infty$. Then, the credible region in (\ref{genericCredibleRegion})
	contains the true parameter $\vbeta_n^0$ with probability approaching $1$.
	\begin{proof}
		We show that $\lim_{n \rightarrow \infty} \bP^\star \{ p_n^{-1} (\widehat{\vbeta}_n - \vbeta_n^0)^\top \mSigma^{-1} (\widehat{\vbeta}_n - \vbeta_n^0) \le C_n /p_n \} = 1$, where $\bP^\star$ denotes the probability measure with respect to the true data generating process as per (B2). By writing $\mathbf{W}_n = \mathbf{X}_n^\top \mathbf{X}_n + \boldsymbol{\Delta}_n$, and the diagonal entries of $\boldsymbol{\Delta}_n$ is non-negative, we have
		$$
		\mathbf{W}_n \ge \mathbf{X}_n^\top \mathbf{X}_n
		$$
		and hence $\mathbf{W}_n^{-1} \le (\mathbf{X}_n^\top \mathbf{X}_n)^{-1}$. Moreover, note that the following identity holds:
		$$
		\widehat{\vbeta}_n - \vbeta_n^0 = \mathbf{W}_n^{-1} \mX_n^\top \vepsilon - \mathbf{W}_n^{-1} \boldsymbol{\Delta}_n \vbeta_n^0.
		$$
		Hence, we may rewrite
		\begin{align*}
			&(\widehat{\vbeta} - \vbeta_n^0)^\top \mSigma^{-1}  (\widehat{\vbeta} - \vbeta_n^0) \\
			&= \widetilde{s}_n^{-1} \{  \mathbf{W}_n^{-1} \mX_n^\top \vepsilon - \mathbf{W}_n^{-1} \boldsymbol{\Delta}_n \vbeta_n^0 \}^\top \mSigma^{-1} \{  \mathbf{W}_n^{-1} \mX_n^\top \vepsilon - \mathbf{W}_n^{-1} \boldsymbol{\Delta}_n \vbeta_n^0 \} \\
			&=  (G_1 - 2 G_2 + G_3)/\widetilde{s}_n, 
		\end{align*}
		where
		\begin{align*}
			&G_1 = \vepsilon^\top \mX_n \mathbf{W}_n^{-1} \mX_n^\top \vepsilon \\
			&G_2 = \vepsilon^\top \mX_n \mathbf{W}_n^{-1} \boldsymbol{\Delta}_n \vbeta_n^0 \\
			&G_3 = \vbeta_n^{0 \, \top} \boldsymbol{\Delta}_n \mathbf{W}_n^{-1} \boldsymbol{\Delta}_n \vbeta_n^0.
		\end{align*}
		Since $\mW_n  \le \mX_n^\top \mX_n + \max_g \delta_g \mI \le \mX_n^\top \mX_n + \frac{\max_g \delta_g}{n d_{\min}} \mX_n^\top \mX_n = \{1+ \frac{\max_g \delta_g}{n d_{\min}} \} \mX_n^\top \mX_n$, hence
		$$
		G_1 \ge \left \{ 1+ \frac{\max_g \delta_g}{n d_{\min}} \right \}^{-1} \vepsilon_n^\top \mX_n ( \mX_n^\top \mX_n  )^{-1} \mX_n^\top \vepsilon_n.
		$$
		Moreover, on the other hand
		$$
		G_1 = \vepsilon_n^\top \mX_n \mathbf{W}_n^{-1} \mX_n^\top \vepsilon_n \le \vepsilon_n^\top \mX_n( \mX_n^\top \mX_n  )^{-1} \mX_n^\top \vepsilon_n
		$$
		Since $1+ \frac{\max_g \delta_g}{n d_{\min}} = 1 + o_p (1)$, we have
		$$
		G_1 /p_n \xrightarrow{\bP^\star} \sigma^{\star \, 2}.
		$$
		We can also bound $G_3$ with
		$$
		G_3 \le \vbeta_n^{0 \, \top} \boldsymbol{\Delta}_n ( \mX_n^\top \mX_n )^{-1} \boldsymbol{\Delta}_n \vbeta_n^0 \le \frac{ \max_g \delta_g^2 \lVert \vbeta_n^0 \rVert^2 }{  n d_{\min} } = o_p(p_n).
		$$
		We upper bound $G_2$ using Cauchy-Schwarz:
		$$
		G_2 \le \sqrt{G_1 G_3} = o_p(p_n).
		$$
		Hence, we have
		$$
		p_n^{-1} (\widehat{\vbeta}_n - \vbeta_n^0)^\top \mSigma^{-1} (\widehat{\vbeta}_n - \vbeta_n^0) \le \frac{G_1 + 2 \lvert G_2 \rvert + G_3}{p_n \widetilde{s_n}} = o_p(1)
		$$
		which completes our proof.
	\end{proof}
\end{lemma} 
\noindent Next, we prove the parameter consistency of the generic posterior expectation. 
\begin{lemma}
	\label{ConsistencyBetahat}
	Suppose (B1) to (B4) holds. Then, $\widehat{\vbeta}_n$ is a consistent estimator of $\vbeta_n^0$ in the sense:
	$$\lVert \widehat{\vbeta}_n - \vbeta_n^0 \rVert^2  \asymp p_n/n.$$
	\begin{proof}
		Recall from Lemma \ref{ContainsTruthProbApproachOne} that
		$$
		\widehat{\vbeta}_n - \vbeta_n^0  = \mW_n^{-1} \mX_n^\top \vepsilon -  \mW_n^{-1} \mDelta_n \vbeta_n^0,
		$$
		We first prove that $\lVert \widehat{\vbeta}_n - \vbeta_n^0 \rVert^2 = O_p(p_n /n )$. In fact, we have
		$$
		\lVert \widehat{\vbeta}_n - \vbeta_n^0 \rVert^2 \le 2 \lVert \mW_n^{-1} \mX_n^\top \vepsilon \rVert^2 + 2 \lVert \mW_n^{-1} \mDelta_n \vbeta_n^0 \rVert^2
		$$
		First observe that
		$$
		\lVert \mW_n^{-1} \mDelta_n \vbeta_n^0 \rVert^2 \le \frac{1}{ \min \text{eigen} (W_n)^2 } \lVert \mDelta_n \vbeta_n^0 \rVert^2 \le \frac{p_n \max_g \delta_g^2 \limsup_{g,n} (u_g^0)^2 }{n^2 d_{\min}^2} = o_p(p_n /n). 
		$$
		Furthermore, we have
		$$
		\lVert \mW_n^{-1} \mX_n^\top \vepsilon \rVert^2 \le \tfrac{1}{n d_{\min}} \vepsilon^\top \mX_n \mW_n^{-1} \mX_n^\top \vepsilon \le \tfrac{1}{n d_{\min}} \vepsilon^\top \mX_n ( \mX_n^\top \mX_n )^{-1} \mX_n^\top \vepsilon.
		$$
		Since $\tfrac{1}{\sigma^{\star \, 2}} \vepsilon^\top \mX_n ( \mX_n^\top \mX_n )^{-1} \mX_n^\top \vepsilon \sim \chi_{p_n}^2$, therefore $\lVert \mW_n^{-1} \mX_n^\top \vepsilon \rVert^2  = O_p(p_n/n)$ and we have proved that $\lVert \widehat{\vbeta}_n - \vbeta_n^0 \rVert^2 = O_p(p/n)$. Next, we show that $\lVert \widehat{\vbeta}_n - \vbeta_n^0 \rVert^2 = \Omega_p(p/n)$ In fact, we have
		$$
		\lVert \widehat{\vbeta}_n - \vbeta_n^0 \rVert \ge \lVert \mW_n^{-1} \mX_n^\top \vepsilon \rVert - \lVert \mW_n^{-1} \mDelta_n \vbeta_n^0 \rVert = \lVert \mW_n^{-1} \mX_n^\top \vepsilon \rVert  \left \{ 1 - \frac{ \lVert \mW_n^{-1} \mDelta_n \vbeta_n^0 \rVert }{  \lVert \mW_n^{-1} \mX_n^\top \vepsilon \rVert } \right \}.
		$$
		Now, as shown previously, $\lVert \mW_n^{-1} \mDelta_n \vbeta_n^0 \rVert = o_p (\sqrt{p_n/n})$. On the other hand,
		\begin{align*}
			\lVert \mW_n^{-1} \mX_n^\top \vepsilon \rVert^2 \ge \frac{1}{n ( d_{\max} + \max_g \delta_g /n ) ( 1 + \max_g \delta_g / (n d_{\min}) ) } \vepsilon^\top \mX_n (\mX_n^\top \mX_n)^{-1} \mX_n^\top \vepsilon
		\end{align*}
		Since $\vepsilon^\top \mX_n (\mX_n^\top \mX_n)^{-1} \mX_n^\top \vepsilon \sim \sigma^{\star \, 2} \chi_{p_n}^2$, we have $\lVert \mW_n^{-1} \mX_n^\top \vepsilon \rVert = \Omega_p (\sqrt{p_n / n})$ and hence we have proved that $\lVert \widehat{\vbeta}_n - \vbeta_n^0 \rVert^2 = \Omega_p (p_n / n)$. Since $\lVert \widehat{\vbeta}_n - \vbeta_n^0 \rVert^2 = \Omega_p(p/n)$ and $\lVert \widehat{\vbeta}_n - \vbeta_n^0 \rVert^2 = \Omega_p (p_n / n)$, we thus have our required result.
	\end{proof}
\end{lemma}
\noindent Between Lemmas \ref{betatildeasymp} and \ref{genericVariableConsistency}, we assume (B5'): There exists a diverging deterministic sequence $b_n^\prime$ such that $a_n b_n^\prime = o(n)$ and $a_n b_n^\prime p_n = o(n)$. Moreover, $\sum_{g \in \sA_n^0} 1/ u_g^0 \le E_0 \sqrt{n / p_n}$ and
$\sum_{g \in \sA_n^0} 1/(u_g^0)^{2} \le E_1 n /(p_n \sqrt{a_n b_n^\prime})$, for some $0 < E_0 < \infty$, $0 < E_1 < \infty$.
\begin{lemma}
	\label{betatildeasymp}
	Assume (B1) to (B4) and (B5') holds. Then, we have
	$$
	( \widetilde{\vbeta}_n - \widehat{\vbeta}_n )^\top \mSigma_n^{-1} ( \widetilde{\vbeta}_n - \widehat{\vbeta}_n ) \asymp a_n p_n.
	$$
	\begin{proof}
		Note that the solution satisfies the equality
		$$
		( \widetilde{\vbeta}_n - \widehat{\vbeta}_n )^\top \mSigma_n^{-1} ( \widetilde{\vbeta}_n - \widehat{\vbeta}_n ) = \min \{ C_n, \widehat{\vbeta}_n^\top  \mSigma_n^{-1} \widehat{\vbeta}_n \}.
		$$
		We examine the divergence rate of $\widehat{\vbeta}_n^\top  \mSigma_n^{-1} \widehat{\vbeta}_n$. In fact,
		\begin{align*}
			\widehat{\vbeta}_n^\top  \mSigma_n^{-1} \widehat{\vbeta}_n  &\ge \frac{n d_{\min} }{ \widetilde{s}_n } \lVert \widehat{\vbeta}_n \rVert^2 \\
			&\ge   \frac{n d_{\min} }{ \widetilde{s}_n } \left \{ \lVert \vbeta^0_n \rVert - \lVert \widehat{\vbeta}_n - \vbeta^0_n \rVert \right \} ^2 \\
			&=  \frac{n d_{\min} }{ \widetilde{s}_n } \lVert \vbeta^0_n \rVert^2 \left \{ 1 - \lVert \widehat{\vbeta}_n - \vbeta^0_n \rVert / \lVert \vbeta^0_n \rVert \right \} ^2 \\
			&\ge  \frac{n d_{\min} }{ \widetilde{s}_n } \max_{g}  (u_g^0)^2 \left \{ 1 - \lVert \widehat{\vbeta}_n - \vbeta^0_n \rVert / \lVert \vbeta^0_n \rVert \right \} ^2 \\
			&= \Omega_p(n)
		\end{align*}
		The first inequality follows from: $\widehat{\vbeta}_n^\top  \widehat{\mSigma}^{-1} \widehat{\vbeta}_n  \ge  \lVert \widehat{\vbeta}_n \rVert^2 \min \text{eigen} (  \mSigma^{-1}  )  $. The second inequality is an application of the reverse triangle inequality. The fourth inequality follows from observing that $\lVert \vbeta^0_n \rVert^2 \ge \max_{g}  (u_g^0)^2 $. The last convergence holds because $\lVert \widehat{\vbeta}_n - \vbeta^0_n \rVert  \asymp  \sqrt{p/n}$,  $\lVert \vbeta^0_n \rVert \rightarrow \infty$, assumption (B4), and the consistency of $\widetilde{s}_n$. On the other hand, by (B5'), we have
		$$\frac{C_n}{n}  \rightarrow 0.$$
		Hence,
		$$
		\lim_{n \rightarrow \infty} \bP^\star \left \{  ( \widetilde{\vbeta}_n - \widehat{\vbeta}_n )^\top \mSigma_n^{-1} ( \widetilde{\vbeta}_n - \widehat{\vbeta}_n ) = C_n \right \} = 1.
		$$
		Now, since $C_n / (p_n a_n) \rightarrow c \in (0, \infty)$, therefore
		$$
		\frac{1}{p_n a_n}( \widetilde{\vbeta}_n - \widehat{\vbeta}_n )^\top \mSigma_n^{-1} ( \widetilde{\vbeta}_n - \widehat{\vbeta}_n ) \xrightarrow{\bP^\star} c \in (0, \infty)
		$$
		and hence we have our required result.
	\end{proof}
\end{lemma}
\begin{lemma}
	\label{genericParameterConsistency}
	Assume (B1) to (B4) and (B5') hold. Then, the generic estimator $\widetilde{\vbeta}_n$ is parameter-consistent for $\vbeta_n^0$. Moreover, 
	$$\lVert \widetilde{\vbeta}_n - \vbeta^0_n \rVert^2 \asymp \frac{a_n p_n}{n}.$$
	\begin{proof}
		We begin by showing that $\lVert \widetilde{\vbeta}_n - \vbeta^0_n \rVert^2 = O_p( a_n p_n /n)$. Observe that
		\begin{align*}
			\lVert \widetilde{\vbeta}_n - \vbeta^0_n \rVert^2 &\le 2\lVert \widetilde{\vbeta}_n - \widehat{\vbeta}_n \rVert^2 + 2\lVert \widehat{\vbeta}_n - \vbeta^0_n \rVert^2.
		\end{align*}
		From Lemma \ref{ConsistencyBetahat}, we have $\lVert \widehat{\vbeta}_n - \vbeta^0_n \rVert^2 = O_p(p_n/n)$. Now for $\lVert \widetilde{\vbeta}_n - \widehat{\vbeta}_n \rVert^2$, we have
		\begin{align*}
			C_n &\ge \min\{ C_n, \widehat{\vbeta}_n^\top \mSigma^{-1} \widehat{\vbeta}_n \} \\
			&= (\widetilde{\vbeta}_n - \widehat{\vbeta}_n)^\top \mSigma^{-1} (\widetilde{\vbeta}_n - \widehat{\vbeta}_n) \\
			&\ge \frac{n d_{\min}  }{ \widetilde{s}_n} \lVert \widetilde{\vbeta}_n - \widehat{\vbeta}_n \rVert^2
		\end{align*}
		Hence
		$$
		\lVert \widetilde{\vbeta}_n - \widehat{\vbeta}_n \rVert^2  \le \frac{\widetilde{s}_n C_n}{n d_{\min}} = O_p( a_n p_n / n). 
		$$
		and consequently $\lVert \widetilde{\vbeta}_n - \widehat{\vbeta}_n \rVert = O_p(\sqrt{ a_n p_n/n } )$. On the other hand
		\begin{align*}
			(\widetilde{\vbeta}_n - \widehat{\vbeta}_n)^\top \widehat{\mSigma}^{-1} (\widetilde{\vbeta}_n - \widehat{\vbeta}_n) \le \frac{n }{ \widetilde{s}_n} \lVert \widetilde{\vbeta}_n - \widehat{\vbeta}_n \rVert^2
		\end{align*}
		And hence
		\begin{align*}
			\lVert \widetilde{\vbeta}_n - \widehat{\vbeta}_n \rVert^2 &\ge \frac{\widetilde{s}_n}{n  } \min\{ C_n, \widehat{\vbeta}_n^\top \widehat{\mSigma}^{-1} \widehat{\vbeta}_n \},
		\end{align*}
		where $\min\{ C_n, \widehat{\vbeta}_n^\top \widehat{\mSigma}^{-1} \widehat{\vbeta}_n \} / (a_n p_n)  \xrightarrow{\bP^\star} c$ by (B5'). This implies that
		$$
		\frac{\lVert \widetilde{\vbeta}_n - \widehat{\vbeta}_n \rVert^2}{ a_n p_n / n } \ge  \frac{\widetilde{s}_n}{ a_n p_n d_{\max}  } \min\{ C_n, \widehat{\vbeta}_n^\top \widehat{\mSigma}^{-1} \widehat{\vbeta}_n \} \xrightarrow{\bP^\star}  \frac{ c \sigma^{\star \, 2} }{ d_{\max} } \in (0, \infty).
		$$
		and hence $\lVert \widetilde{\vbeta}_n - \widehat{\vbeta}_n \rVert^2 = \Omega_p(a_n p_n / n)$. By noting from previous segment of this proof that $\lVert \widetilde{\vbeta}_n - \widehat{\vbeta}_n \rVert^2 = O_p(a_n p_n / n)$, we thus have $\lVert \widetilde{\vbeta}_n - \widehat{\vbeta}_n \rVert^2 \asymp a_n p_n / n$. By reverse triangle inequality, we have
		\begin{align*}
			\lVert \widetilde{\vbeta}_n - \vbeta^0_n \rVert &\ge \left \lvert  \lVert \widetilde{\vbeta}_n - \widehat{\vbeta}_n \rVert - \lVert \widehat{\vbeta}_n - \vbeta^0_n \rVert \right \rvert \\
			&= \lVert \widetilde{\vbeta}_n - \widehat{\vbeta}_n \rVert \left \lvert 1 - \frac{ \lVert \widehat{\vbeta}_n - \vbeta^0_n \rVert }{ \lVert \widetilde{\vbeta}_n - \widehat{\vbeta}_n \rVert }\right \rvert.
		\end{align*} 
		Since $ \lVert \widehat{\vbeta}_n - \vbeta^0_n \rVert^2  \asymp p/n$, $\lVert \widetilde{\vbeta}_n - \widehat{\vbeta}_n \rVert^2 \asymp a_n p_n /n$, and by taking note of (B5'), we have our required result.
	\end{proof}
\end{lemma}
\section{Variable selection consistency of generic GroupPenCR estimator}
\noindent In the next segment of results, we establish the variable selection consistency of the generic \texttt{GroupPenCR} estimator $\widetilde{\vbeta}_n$. To enhance the clarity of the results, we denote the set of selected groups using $\widetilde{\vbeta}_n$ by:
$$
\sA_n = \{ g \in \sG_n \, : \, \widetilde{u}_g \neq 0 \},
$$
where $\widetilde{u}_g = \lVert \widetilde{\vbeta}_{g;n} \rVert$. We also denote the objective function of the grouped penalized credible region estimator as
$$
K(\vbeta) = \sum_{g=1}^G \sqrt{p_g} \frac{u_g}{ \widehat{u}_g^2},
$$
where $u_g := u_g (\vbeta_{g;n}) = \lVert \vbeta_{g;n} \rVert$. The proofs in this section are extensions of those presented in \cite{Zhang2018} to the grouped variable selection setting.
\begin{lemma}
	\label{nonSignificantLimit}
	Assume (B1) to (B4) and (B5') hold. Then, $\sqrt{n/p_n} \left \lVert \widetilde{\vbeta}_{g;n} - \widehat{\vbeta}_{g;n} \right \rVert = O_p(1)$ for all $g \notin \sA_n^0$.
	\begin{proof}
		We use a proof-by-contradiction approach. Suppose $l \notin \sA_n^0$ and $(n/p_n) \lVert \widetilde{\vbeta}_{l;n} - \widehat{\vbeta}_{l;n} \rVert^2 \rightarrow \infty$. Then $K(\widetilde{\vbeta}) \ge \sqrt{p_l}  (n/p_n) \widetilde{u}_l / \{ (n/p_n) \widehat{u}_l^2 \}$. We analyse the numerator and denominator of the lower bound. Note that $ \lVert \widehat{\vbeta}_n - \vbeta^0_n \rVert^2 \asymp p_n /n$, we have
		$(n/p_n) \widehat{u}_l^2$ is bounded away from $\infty$ in probability. Also, we compute a lower bound for
		\begin{align*}
			\widetilde{u}_l &= \lVert \widetilde{\vbeta}_{l;n} \rVert \\
			&\ge \lvert \lVert \widetilde{\vbeta}_{l;n} - \widehat{\vbeta}_{l;n} \rVert - \lVert \widehat{\vbeta}_{l;n} - \vbeta_{l;n}^0 \rVert \rvert \\
			&= \lVert \widetilde{\vbeta}_{l;n} - \widehat{\vbeta}_{l;n} \rVert \left \lvert 1 - \frac{\sqrt{n/p} \lVert \widehat{\vbeta}_{l;n} - \vbeta_{l;n}^0 \rVert }{ \sqrt{n/p} \lVert \widetilde{\vbeta}_{l;n} - \widehat{\vbeta}_{l;n} \rVert  } \right \rvert
		\end{align*}
		Since $\sqrt{n/p_n} \lVert \widehat{\vbeta}_{l;n} - \vbeta_{l;n}^0 \rVert = O_p(1)$ and $\sqrt{n/p_n} \lVert \widetilde{\vbeta}_{l;n} - \widehat{\vbeta}_{l;n} \rVert \rightarrow \infty$, we have $\sqrt{n/p_n} \widetilde{u}_l \rightarrow \infty$ and hence $\sqrt{p_n /n } K(\widetilde{\vbeta})  \rightarrow \infty$, for each $g = 1, \ldots, G_n$. Now, consider the subset within the generic credible region $\overline{\sS}_n = \{ \vbeta \, : \, u_g = 0 \; \forall \; g \notin \sA_n^0, \, \vbeta \in \sS  \} \subseteq \sS$ and the corresponding minimizer $\breve{\vbeta}_n$ in this restricted set. Following Lemma \ref{ContainsTruthProbApproachOne}, because $\vbeta_n^0 \in \mathcal{S}_n$ with probability approaching 1 and $\vbeta_n^0$ satisfies the constraints of $\overline{\mathcal{S}}_n$, we know that $\overline{\mathcal{S}}_n$ is non-empty with probability approaching 1. Using similar arguments to Lemma \ref{ConsistencyBetahat} and Lemma \ref{betatildeasymp}, we have $\{n/(p_n a_n)\} \lVert \breve{\vbeta}_{g;n} - \widehat{\vbeta}_{g;n} \rVert^2 = O_p(1)$ and $(n/p_n) \lVert \widehat{\vbeta}_{g;n} - \vbeta_{g;n}^0 \rVert^2 = O_p(1)$. Let $T_n = \sum_{g \in \sA_n^0}^{G_n} \sqrt{p_g} / u_g^0$. We bound
		\begin{align*}
			\lvert K ( \breve{\vbeta}_n ) - T_n \rvert &\le \sum_{g \in \sA_n^0} \sqrt{p_g} \left \lvert \frac{ \breve{u}_g }{ \widehat{u}_g^2 } - \frac{1}{u_g^0} \right \rvert \\
			&= \sum_{g \in \sA_n^0} \sqrt{p_g} \left \lvert \frac{ (\breve{u}_g - \widehat{u}_g) u_g^0 + \widehat{u}_g( u_g^0 - \widehat{u}_g ) }{\widehat{u}_g^2 u_g^0 }  \right \rvert \\
			&\le \sum_{g \in \sA_n^0} \frac{\sqrt{p_g}}{ \widehat{u}_g^2 } \lVert \breve{\vbeta}_{g;n} - \widehat{\vbeta}_{g;n} \rVert + \sum_{g \in \sA_n^0} \frac{\sqrt{p_g}}{ \widehat{u}_g u_g^0 } \lVert \widehat{\vbeta}_{g;n} - \vbeta_{g;n}^0 \rVert, 
		\end{align*}
		where the last inequality follows from the reverse-triangle inequality. 
		
		\noindent Define the event $\sE_n = \left \{ \max_{g \in \sA_n^0} \lVert \widehat{\vbeta}_{g;n} - \vbeta_{g;n}^0 \rVert \le \tfrac{1}{2}  \min_{g \in \sA_n^0} u_g^0  \right \}$. Following (B5'), we have
		$$
		\frac{1}{\min_{g \in \sA_n^0} (u_g^0)^2 } \le \sum_{g \in \sA_n^0} \frac{1}{ (u_g^0)^2 } \le \frac{E_1 n}{ p_n \sqrt{a_n b_n^\prime} }
		$$
		and hence by rearranging we have
		$$
		\min_{g \in \sA_n^0}  u_g^0 \ge \sqrt{ \frac{p_n \sqrt{a_n b_n^\prime}}{ n E_1} }
		$$
		or $\min_{g \in \sA_n^0}  u_g^0  = \Omega \left ( \sqrt{ \frac{p_n \sqrt{a_n b_n^\prime}}{ n } } \right )$. On the other hand, following Lemma \ref{ConsistencyBetahat}, we have $ \max_{g \in \sA_n^0} \lVert \widehat{\vbeta}_{g;n} - \vbeta_{g;n}^0 \rVert = O_p(\sqrt{p_n /n}) = o_p \left (   \sqrt{ \frac{p_n \sqrt{a_n b_n^\prime}}{ n } }  \right )$. Thus the probability of $\bP^\star(\sE_n) \rightarrow 1$. Now, conditioning on the event $\sE_n$, we have
		\begin{align*}
			\widehat{u}_g &\ge u_g^0 - \lvert \widehat{u}_g - u_g^0 \rvert \\
			&\ge u_g^0 - \lVert \widehat{\vbeta}_{g;n} - \vbeta_{g;n}^0 \rVert \\
			&\ge u_g^0 - \tfrac{1}{2} \min_g u_g^0 = \tfrac{1}{2} u_g^0
		\end{align*}
		and by combining with Lemma \ref{genericParameterConsistency} and (B5'), we have
		$$
		\sqrt{\frac{p_n}{n}} \sum_{g \in \sA_n^0} \frac{\sqrt{p_g}}{ \widehat{u}_g^2 } \lVert \breve{\vbeta}_{g;n} - \widehat{\vbeta}_{g;n} \rVert \le 	\sqrt{\frac{p_n p_{\max} }{n}} \lVert \breve{\vbeta}_{n} - \widehat{\vbeta}_{n} \rVert \sum_{g \in \sA_n^0} \frac{4}{ (u_g^0)^2 } = o_p(1)
		$$
		and
		$$
		\sqrt{\frac{p_n}{n}} \sum_{g \in \sA_n^0} \frac{\sqrt{p_g}}{ \widehat{u}_g u_g^0 }  \lVert \widehat{\vbeta}_{g;n} - \vbeta_{g;n}^0 \rVert \le \sqrt{\frac{p_n p_{\max}}{n}}   \lVert \breve{\vbeta}_{n} - \widehat{\vbeta}_{n} \rVert \sum_{g \in \sA_n^0} \frac{2}{ (u_g^0)^2 } = o_p(1).
		$$
		Moreover, we can also show that $\sqrt{p_n / n} T_n = O(1)$ from (B5'). And hence, conditioned on $\sE_n$, $\sqrt{p_n/n} K ( \breve{\vbeta}_n ) = O_p(1)$. Now, clearly due to the definition of $\widetilde{\vbeta}_n$ and $\breve{\vbeta}_n$, they must satisfy the inequality $K( \breve{\vbeta}_n ) \ge K( \widetilde{\vbeta}_n )$ for all $n$. However, since $\sqrt{p_n/n} K ( \breve{\vbeta}_n ) = O_p(1)$ and $\sqrt{p_n/n} K ( \widetilde{\vbeta}_n ) \xrightarrow{\bP^\star } \infty$, we have
		$$
		\bP^\star \left \{  \sqrt{p_n / n} K( \breve{\vbeta}_n ) \ge \sqrt{p_n / n} K( \widetilde{\vbeta}_n ) \right \} \rightarrow 0
		$$
		which leads us to a contradiction.
	\end{proof}
\end{lemma}
\begin{lemma}
	\label{genericVariableConsistency}
	Assume (B1) to (B4) and (B5') holds. Then, $\lim_{n \rightarrow \infty} \bP^\star ( \sA_n = \sA_n^0 ) = 1$.
	\begin{proof}
		\noindent Let $\sR_{-g} = \{ \vbeta \in \bR^p \, : \, u_g = 0 \}$.
		We need only to show for any $g \in \sA_n^0$, we have $\bP^\star ( g \notin \sA_n) \rightarrow 0$ (false negative rate go to $0$) and for any $g \notin \sA_n^0$, we have $\bP^\star ( g \in \sA_n) \rightarrow 0$ (false positive rate go to $0$). For any $g \in \sA_n^0$, consider the the subset $\sR_{-g} \subset \bR^{p_n}$. Now,
		\begin{align*}
			\bP^\star \left ( g \notin \sA_n \right ) \le \bP^\star \left ( \sR_{-g} \cap \sS_n \neq \emptyset \right ) = 1 - \bP^\star \left ( \min_{\vbeta \in \sR_{-g}} (\vbeta - \widehat{\vbeta}_n )^\top \widehat{\mSigma}^{-1} (\vbeta - \widehat{\vbeta}_n ) >  C_n \right ).
		\end{align*}
		For any $\overline{\vbeta} \in \sR_{-g}$, we have the lower bound
		$$
		(\overline{\vbeta} - \widehat{\vbeta}_n )^\top \widehat{\mSigma}^{-1} ( \overline{\vbeta} - \widehat{\vbeta}_n ) \ge \frac{n d_{\min} }{ \widetilde{s}_n} \lVert \widehat{\vbeta}_{g;n} \rVert^2
		$$
		Hence
		$$
		\bP^\star \left ( \min_{\vbeta \in \sR_{-g}} (\vbeta - \widehat{\vbeta}_n )^\top \widehat{\mSigma}^{-1} (\vbeta - \widehat{\vbeta}_n ) >  C_n \right ) \ge \bP^\star \left \{ \lVert \widehat{\vbeta}_{g;n} \rVert^2 >  \frac{C_n \widetilde{s}_n}{n d_{\min}}  \right \} \rightarrow 1,
		$$
		where the last convergence follows from noting that $\lVert \widehat{\vbeta}_{g;n} \rVert^2 = (u_g^0)^2 + O_p(p/n)$, $u_g^{0} \neq 0$, $C_n \asymp a_n p_n$, and $a_n p_n / n \rightarrow 0$. Hence, $\bP^\star \left \{ \min_{\vbeta \in \sR_{-g}} (\vbeta - \widehat{\vbeta}_n )^\top \widehat{\mSigma}^{-1} (\vbeta - \widehat{\vbeta}_n ) >  C_n \right \} \rightarrow 1$ and consequently $\bP^\star \left ( g \notin \sA_n \right ) \rightarrow 0$. \\
		\noindent Next, we show that for any $g \notin \sA_n^0$, we have $\bP^\star ( g \in \sA_n) \rightarrow 0$. We adopt a proof-by-contradiction approach, i.e., we assume there exists a group such that the probability of selecting it approaches non-zero but it is not a truly important group. Without loss of generality, assume that the first $q_n$ groups are truly important, i.e., $\sA_n^0 = \{1 , \ldots, q_n  \}$, $\bP^\star \{ \widetilde{u}_G \neq 0 \} \rightarrow v > 0$, and that we rearrange the order of $\vx_1$ such that $(1,p_1) = \argmax_{g \in \sA_n^0, j \in [p_g]} \lvert \widetilde{\beta}_{g,j;n} - \widehat{\beta}_{g,j;n} \rvert$, where $\beta_{g,j;n}$ denotes the $j$-th entry of $\vbeta_{g;n}$. Here, our assumptions imply that index $G_n$ is reserved for that particular group with selection probability approaching a non-zero constant $v$ but is not a truly important group. By noting that $(\widetilde{\vbeta}_n - \widehat{\vbeta}_n)^\top \mSigma^{-1} (\widetilde{\vbeta}_n - \widehat{\vbeta}_n) = \min \left \{ C_n, \widehat{\vbeta}_n^\top \mSigma^{-1} \widehat{\vbeta}_n \right \}$, thus $\widetilde{\beta}_{1,p_1;n}$ can be expressed as an implicit function of the remaining entries of $\widetilde{\vbeta}_n$. Consider the events $\sK_n = \{ \widetilde{u}_G \neq 0 \}$ and
		$$
		\sB_n = \left \{ \sqrt{p_G} \frac{ \widetilde{\beta}_{G, j^\prime;n} }{  \widetilde{u}_{G} \widehat{u}_G^2  } + \sqrt{p_1} \frac{ \widetilde{\beta}_{1,p_1;n} }{  \widetilde{u}_{1} \widehat{u}_1^2  } \frac{\partial \widetilde{\beta}_{1,p_1;n} }{\partial \widetilde{\beta}_{G, j^\prime;n} } = 0 \right \},
		$$
		for some $j^\prime \in [p_{G_n}]$. Then, by KKT conditions, we have $\sK_n \subseteq \sB_n$. Since,
		$$
		\bP^\star( \sK_n ) = \bP^\star ( \sK_n, \sB_n ) = \sP^\star ( \sB_n \mid \sK_n) \sP^\star (\sK_n)
		$$
		it is necessary that $\limsup_{n \rightarrow \infty} \sP^\star ( \sB_n \mid \sK_n) = 1$. It remains for us to show that $\limsup_{n \rightarrow \infty} \sP^\star ( \sB_n \mid \sK_n) < 1$.
		Conditioned on $\mathcal{K}_n$, let $j' \in \{1, ..., p_{G_n}\}$ be the index of the coordinate with the maximum absolute value in group $G_n$, which must satisfy $\widetilde{\beta}_{G_n, j'; n} \ne 0$. By Lemma \ref{nonSignificantLimit}, we established $(n/p_n)\widehat{u}_{G_n}^2 = O_p(1)$. Since the maximum coordinate satisfies $|\widetilde{\beta}_{G_n, j'; n}| / \widetilde{u}_{G_n} \ge 1/\sqrt{p_{max}}$, the magnitude of the first KKT term is strictly bounded below in probability:$$\left| \sqrt{p_{G_n}} \frac{\widetilde{\beta}_{G_n, j'; n}}{\widetilde{u}_{G_n} \widehat{u}_{G_n}^2} \right| \ge \frac{1}{\sqrt{p_{max}}} \frac{1}{\widehat{u}_{G_n}^2} = \Omega_p\left(\frac{n}{p_n}\right).$$On the other hand, due to parameter consistency, $\widetilde{u}_1 \to u_1^0 > 0$, we have $\sqrt{p_1} \widetilde{\beta}_{1, p_1; n} / (\widetilde{u}_1 \widehat{u}_1^2) = O_p(1)$. Using implicit differentiation of $(\widetilde{\vbeta}_n - \widehat{\vbeta}_n)^\top \mSigma_n^{-1} (\widetilde{\vbeta}_n - \widehat{\vbeta}_n) = \min \left \{ C_n, \widehat{\vbeta}_n^\top \mSigma_n^{-1} \widehat{\vbeta}_n \right \}$, we obtain:
		\begin{equation}
			\label{KeyDerivativeToBound}
			\frac{\partial \widetilde{\beta}_{1,p_1;n}}{\partial \widetilde{\beta}_{G_n,j';n}} = - \frac{ [\mV_n]_{G_n, j'} }{ [\mV_n]_{1, p_1} },
		\end{equation}
		where $\mV_n = \mSigma_n^{-1} (\widetilde{\vbeta}_n - \widehat{\vbeta}_n)$.
		To rigorously bound this derivative, we establish the asymptotic orders of the numerator and denominator in four steps:
		\begin{enumerate}
			\item Dominance by active set:  We first show that the contribution from $\sum_{g \in \sA_n^0} \lVert \widetilde{\vbeta}_{g;n} - \widehat{\vbeta}_{g;n} \rVert^2$ dominates $\lVert \widetilde{\vbeta}_n - \widehat{\vbeta}_n \rVert^2$. To do so, we recall from Lemma \ref{genericParameterConsistency} that $\lVert \widetilde{\vbeta}_n - \widehat{\vbeta}_n \rVert^2 = \Omega_p ( a_n p_n /n )$. It remains for us to show that $\sum_{g \notin \sA_n^0} \lVert \widetilde{\vbeta}_g - \widehat{\vbeta}_g \rVert^2 = o_p( a_n p_n /n) $. In fact,
			$$
			\sum_{g \notin \sA_n^0} \lVert \widetilde{\vbeta}_{g;n} - \widehat{\vbeta}_{g;n} \rVert^2 \le 2 \sum_{g \notin \sA_n^0} \widetilde{u}_{g;n}^2 + 2 \sum_{g \notin \sA_n^0} \widehat{u}_{g;n}^2.
			$$
			Now, $\sum_{g \notin \sA_n^0} \widehat{u}_{g;n}^2 = \sum_{g \notin \sA_n^0} \lVert  \widehat{\vbeta}_{g;n} - \vbeta_{g;n}^0 \rVert^2 \le \lVert \widetilde{\vbeta}_n - \widehat{\vbeta}_n \rVert^2 = O_p(p_n / n)$. Moreover, $\sum_{g \notin \sA_n^0} \widetilde{u}_{g;n} \le \sum_{g \notin \sA_n^0} \widehat{u}_{g;n}^2 \sum_{g \notin \sA_n^0} \tfrac{\widetilde{u}_{g;n}}{ \widehat{u}_{g;n}^2 } \le \sum_{g \notin \sA_n^0} \widehat{u}_{g;n}^2  \sum_{g =1 }^{G_n} \tfrac{\widetilde{u}_{g;n}}{ \widehat{u}_{g;n}^2 } = O_p(p_n/n) O_p(\sqrt{n/p_n}) = O_p(\sqrt{p_n / n})$, where the second last stochastic convergence equality of $\sum_{g =1 }^{G_n} \tfrac{\widetilde{u}_{g;n}}{ \widehat{u}_{g;n}^2 }$ follows from Lemma \ref{nonSignificantLimit}. Hence, we have shown that $\sum_{g \notin \sA_n^0} \lVert \widetilde{\vbeta}_{g;n} - \widehat{\vbeta}_{g;n} \rVert^2 = O_p(p_n/n) = o_p( a_n p_n / n )$. Consequently, we must have $\sum_{g \in \sA_n^0} \lVert \widetilde{\vbeta}_{g;n} - \widehat{\vbeta}_{g;n} \rVert^2 = \Omega_p (a_n p_n /n)$.
			\item Bound for denominator of \eqref{KeyDerivativeToBound}: Since $\sum_{g \in \sA_n^0} \lVert \widetilde{\vbeta}_{g;n} - \widehat{\vbeta}_{g;n} \rVert^2 = \Omega_p (a_n p_n /n)$ and there are at most $q_n p_{\max}$ coordinates from the active set, we thus have $\lvert \widetilde{\beta}_{1,p_1;n} - \widehat{\beta}_{1,p_1;n}\rvert = \Omega_p (\sqrt{a_n p_n / (n q_n )})$. Moreover, by (B3), the posterior covariance has an asymptotic rate of $\mSigma_n^{-1} \asymp n \mI$. Consequently, the denominator can be lower bounded as:
			$$\left \lvert [\mV_n]_{1, p_1} \right\rvert \ge n d_{\min} \lvert \widetilde{\beta}_{1,p_1;n} - \widehat{\beta}_{1,p_1;n}\rvert  = \Omega_p (\sqrt{n a_n p_n /  q_n }).$$
			\item Bound for numerator of \eqref{KeyDerivativeToBound}: By Lemma \ref{genericParameterConsistency} that $ \lVert \widetilde{\vbeta}_n - \widehat{\vbeta}_n \rVert = O_p(\sqrt{a_n p_n / n})$. Hence, we have $$ \left \lvert [\mV_n]_{G_n, j'} \right \rvert \le \lVert \mV_n \rVert  = O_p\left( n \sqrt{\frac{a_n p_n}{n}} \right) = O_p(\sqrt{n a_n p_n}).$$
			\item Final derivative rate: Taking the absolute ratio of the bounded numerator to the bounded denominator yields:$$\left| \frac{\partial \tilde{\beta}_{1,p_1;n}}{\partial \tilde{\beta}_{G_n,j';n}} \right| = \frac{O_p(\sqrt{n a_n p_n})}{\Omega_p(\sqrt{n a_n p_n / q_n})} = O_p(\sqrt{q_n}) \le O_p(\sqrt{p_n}).$$
			Since an $\Omega_p(n/p_n)$ term cannot equal an $O_p(\sqrt{p_n})$ for all plausible regimes of $p_n = o (n)$ term as $n \to \infty$ (take for example, $p_n \asymp n^{1/2}$), we must have $\limsup_{n\to\infty} \mathbb{P}^*(\mathcal{B}_n | \mathcal{K}_n) = 0$, which completes the contradiction.
		\end{enumerate}
	\end{proof}
\end{lemma}

\section{Properties of CAVI posterior parameters}
In this section, we study the properties of the \texttt{VGPenCR} estimator. The proofs in this section do not bear any similarity to those in existing works.
\begin{lemma}
	\label{ConvergenceForSigmaSqMbg}
	Assume (B1) to (B5) holds. Following the initialisations in (\#) of the main manuscript, we have $\lvert 1/m_{1/\sigma^2}^{(t)} - (\sigma^{\star})^2 \rvert = O_p(1)$, and $\max_g m_{b_g}^{(t)}/\tau = O_p \left ( \sqrt{ a_n^{(0)} } \right )$ for all $t \ge 0$. In particular, for all $t \ge 2$, we have $\lvert 1/m_{1/\sigma^2}^{(t)} -  (\sigma^{\star})^2 \rvert = O_p( \max\{ 1 / \sqrt{n}, p/n \} )$.
	\begin{proof}
		For $t = 0$,
		\begin{align*}
			\left \lvert 1/ m_{1/\sigma^2}^{(0)} - (\sigma^\star)^{2}  \right \rvert	&= 		\left	\lvert \frac{ \lVert \vy - \mX_n \widehat{\vbeta}_{LS} \rVert^2 / (n - p)}{ 1 + \xi_n \lVert \vy - \mX_n \widehat{\vbeta}_{LS} \rVert^2/(n - p) }  - (\sigma^\star)^{2}  \right \rvert \\
			&= 		\left	\lvert \frac{ \lVert \vy - \mX_n \widehat{\vbeta}_{LS} \rVert^2 / (n - p) - (\sigma^\star)^{2} -  (\sigma^\star)^{2}  \xi_n \lVert \vy - \mX_n \widehat{\vbeta}_{LS} \rVert^2/(n - p)  }{ 1 + \xi_n \lVert \vy - \mX_n \widehat{\vbeta}_{LS} \rVert^2/ (n - p) }   \right \rvert \\
			&\le \left	\lvert \frac{ \lVert \vy - \mX_n \widehat{\vbeta}_{LS} \rVert^2 / (n - p) - (\sigma^\star)^{2}   }{ 1 + \xi_n \lVert \vy - \mX_n \widehat{\vbeta}_{LS} \rVert^2/(n - p) }  \right \rvert + (\sigma^\star)^{2} \\
			&= O_p(\sqrt{1/n}) + O_p(1) = O_p(1).
		\end{align*}
		Note that the second last equality follows from an application of Chebychev's inequality to show that
		$$
		\left	\lvert \lVert \vy - \mX_n \widehat{\vbeta}_{LS} \rVert^2 / (n-p) - (\sigma^\star)^{2}    \right \rvert =  O_p(1/\sqrt{n}).
		$$
		\noindent Hence, we also have $1/ m_{1/\sigma^2}^{(0)} = O_p(1)$. 
		Since $\limsup_g m_{b_g}^{(0)} < B$, we have $\max_g m_{b_g}^{(0)} / \tau = O_p( \sqrt{a_n^{(0)}} )$. In fact, for any $t \ge 0$, we have
		$$
		\frac{\max_g m_{b_g}^{(t+1)}}{\tau} \le \frac{(p_g + 1) (1 + \max_g m_{b_g}^{(t)}) }{2 \tau} =  O_p( \sqrt{a_n^{(0)}} ).
		$$
		By (B5), $ a_n^{(0)} = o(n)$ and hence, for any $t \ge 1$, the VB posterior mean
		$$
		\vmu_{\vbeta}^{(t)} = \left \{ \mX_n^\top \mX_n + \mM_\tau^{(t-1)} \right \}^{-1} \mX_n^\top \vy
		$$
		satisfies the same asymptotic conditions as the generic posterior expectation in equation (\ref{eqn::genericBetaHat}). By Lemma \ref{ConsistencyBetahat}, for any $t \ge 1$, we have 
		$$
		\lVert \vmu_{\vbeta}^{(t)} - \widehat{\vbeta}_{LS} \rVert^2 \asymp p/n.
		$$
		Furthermore, for any $t \ge 1$, we have
		\begin{align*}
			&\frac{1}{m_{1/\sigma^2}^{(t)}} - (\sigma^\star)^2 \\
			&= \frac{2s + \left [ \vmu_{\vbeta}^{(t-1) \top} \mM_\tau^{(t-1)}  \vmu_{\vbeta}^{(t-1)} + p/m_{1/\sigma^2}^{(t-1)}  + \lVert \vy - \mX_n \vmu_{\vbeta}^{(t-1)} \rVert^2 \right ] }{2r + n + p} - (\sigma^\star)^2\\
			&= \frac{2s + p/m_{1/\sigma^2}^{(t-1)}}{2r + n + p} + \frac{ \vmu_{\vbeta}^{(t-1) \top} \mM_\tau^{(t-1)}  \vmu_{\vbeta}^{(t-1)} }{2r + n +p} + \frac{ \lVert \vy - \mX_n \vmu_{\vbeta}^{(t-1)} \rVert^2 }{2r + n +p} - (\sigma^\star)^2 \\
			&\le \frac{2s + p/m_{1/\sigma^2}^{(t-1)}}{2r + n + p} + \frac{\max_{g} m_{b_g}^{(t-1)} \lVert \vmu_{\vbeta}^{(t-1)}  \rVert^2 }{n \tau} + \frac{ \lVert \vy - \mX_n \widehat{\vbeta}_{LS} \rVert^2 }{n} \\
			&+ \frac{ ( \vmu_{\vbeta}^{(t-1)}  - \widehat{\vbeta}_{LS})^\top \mX_n^\top \mX_n ( \vmu_{\vbeta}^{(t-1)}  - \widehat{\vbeta}_{LS}) }{n} - (\sigma^\star)^2 \\
			&\le \frac{2s + p/m_{1/\sigma^2}^{(t-1)}}{2r + n + p} + \frac{p \max_{g} m_{b_g}^{(t-1)} \limsup_{n,g} (u_g^0)^2 }{n \tau} + \frac{\sigma^{\star \, 2} \chi_{n-p}^2 }{n - p} - (\sigma^\star)^2 \\
			&+ d_{\max} \lVert \vmu_{\vbeta}^{(t-1)}  - \widehat{\vbeta}_{LS} \rVert^2, \\
			&\le \frac{2s + p/m_{1/\sigma^2}^{(t-1)}}{2r + n + p} + \frac{p \max_{g} m_{b_g}^{(t-1)} \limsup_{n,g} (u_g^0)^2 }{n \tau} + \left \lvert  \frac{\sigma^{\star \, 2} \chi_{n-p}^2 }{n - p} - (\sigma^\star)^2 \right \rvert \\
			&+ d_{\max} \lVert \vmu_{\vbeta}^{(t-1)}  - \widehat{\vbeta}_{LS} \rVert^2
		\end{align*}
		\noindent Hence for $t = 1$, we have $\lvert 1/m_{1/\sigma^2}^{(1)} - (\sigma^\star)^2 \rvert = O_p( \lVert \vmu_{\vbeta}^{(0)}  - \widehat{\vbeta}_{LS} \rVert^2) = o_p( n/(a_n^{(0)} p_n) )$. For $t \ge 2$,  we have $\lvert 1/m_{1/\sigma^2}^{(t)} - (\sigma^\star)^2 \rvert = O_p \left ( \left \lvert  \sigma^{\star \, 2} \chi_{n-p}^2/(n-p)  - (\sigma^\star)^2 \right \rvert \right ) + O_p(\lVert \vmu_{\vbeta}^{(t-1)}  - \widehat{\vbeta}_{LS} \rVert^2) = O_p( \max\{ 1 / \sqrt{n}, p/n \} ) $.
	\end{proof}
\end{lemma}

\begin{lemma}
	\label{ConvergenceForMuBetaMv}
	Assume (B1) to (B5) holds. Following the initialisations in (\#) of the main manuscript, we have $\lVert 	\widetilde{\vbeta}_{CAVI}^{(1)} - \vmu_{\vbeta}^{(1)} \rVert = o_p( 1  )$ and $\lVert 	\widetilde{\vbeta}_{CAVI}^{(t)} - \vmu_{\vbeta}^{(t)} \rVert \asymp  \sqrt{ a_n^{(t)} p_n / n } $ for all $t \ge 2$.
	\begin{proof}
		We observe that
		\begin{align*}
			\lVert 	\widetilde{\vbeta}_{CAVI}^{(1)} - \vmu_{\vbeta}^{(1)} \rVert^2	&\le \frac{ (1/m_{1/\sigma^2}^{(1)}) C_n^{(1)} }{n d_{\min}} \\
			&\le \frac{ \lvert 1/m_{1/\sigma^2}^{(1)} - (\sigma^\star)^2 \rvert C_n^{(1)} }{n d_{\min}} +  \frac{ (\sigma^\star)^2 C_n^{(1)} }{n d_{\min}} = O_p \left ( \frac{a_n^{(1)}}{a_n^{(0)}} \right ) = o_p(1),
		\end{align*}
		where the second last equality follows by noting that $C_n^{(1)} \asymp a_n^{(1)} p_n$ and $\lvert 1/m_{1/\sigma^2}^{(1)} - (\sigma^\star)^2 \rvert  = o_p( n/(a_n^{(0)} p_n) )$. For $t \ge 2$, we have shown that $\lvert 1/m_{1/\sigma^2}^{(t)} - (\sigma^\star)^2 \rvert = O_p( \max\{ 1 / \sqrt{n}, p/n \} )$. Hence, by Lemma \ref{genericParameterConsistency}, $\lVert 	\widetilde{\vbeta}_{CAVI}^{(t)} - \vmu_{\vbeta}^{(t)} \rVert \asymp \sqrt{C_n^{(t)} / n} \asymp  \sqrt{ a_n^{(t)} p_n / n } $ for all $t \ge 2$.
	\end{proof}
\end{lemma}

\subsection{Proof of Theorem 4.1}
Following Lemma \ref{ConvergenceForMuBetaMv}, for any $t \ge 2$, the CAVI posterior mean $\vmu_{\vbeta}^{(t)}$ satisfies the conditions of a generic posterior expectation in (\ref{eqn::genericBetaHat}) and $\mSigma_{\vbeta}$ satisfies the conditions of a generic posterior covariance in (\ref{eqn::genericSigma}). Since our proposed CAVI-based estimator is
$$
\widetilde{\vbeta}_{CAVI}^{(t)} = \argmin_{\vbeta \in S_{CAVI}^{(t)}} K(\vbeta)
$$
where
$$
S_{CAVI}^{(t)} = \{ \vbeta \, : \, (\vbeta - \vmu_\vbeta^{(t)})^\top \mSigma_{\vbeta}^{^{(t)} \, -1} (\vbeta - \vmu_\vbeta^{(t)}) \le C_n^{(t)} \}
$$
and $C_n^{(t)} \asymp a_n^{(t)} p_n$, consequently $\widetilde{\vbeta}_{CAVI}^{(t)}$ satisfies the conditions of a generic \texttt{GroupPenCR} estimator in (\ref{genericGroupPenCREstimatorForm}). Hence, $\lVert \widetilde{\vbeta}_{CAVI}^{(t)} - \vbeta^0_n \rVert \asymp \sqrt{a_n^{(t)} p_n /n}$ by Lemma \ref{genericParameterConsistency} and thus $\widetilde{\vbeta}_{CAVI}^{(t)}$ is parameter-consistent for $\vbeta^0_n$. Furthermore, by Lemma \ref{genericVariableConsistency}, we have $\bP \left \{ \sA_{CAVI}^{(t)} = \sA_{n}^0 \right \} \rightarrow 1$ and thus $\widetilde{\vbeta}_{CAVI}^{(t)}$ is variable selection-consistent for $\vbeta^0$.

\subsection{Proof of Theorem 2}
From Theorem 1, we have
$$
\frac{ \lVert \widetilde{\vbeta}_{CAVI}^{(t+1)} - \vbeta^0_n \rVert }{ \lVert \widetilde{\vbeta}_{CAVI}^{(t)} - \vbeta^0_n \rVert } \le G_n b_n^{- \frac{1}{2(t+1)(t+2)}} \xrightarrow{\bP^\star} 0,
$$
where the last convergence holds from (B5) and noting that $\lVert \widetilde{\vbeta}_{CAVI}^{(t+1)} - \vbeta^0_n \rVert \asymp \sqrt{a_n^{(t+1)} p_n /n}$ and $\lVert \widetilde{\vbeta}_{CAVI}^{(t)} - \vbeta^0_n \rVert \asymp \sqrt{a_n^{(t)} p_n /n}$ implies $G_n = O_p(1)$.

\section{Derivation of mean-field approximate posteriors}
\subsection{Derivation for $q(\vbeta)$}
{\small
	\begin{align*}
		q(\vbeta) &\propto \exp \left[  \bE_q \left [ \log p( \vy \mid \vbeta, \sigma^2 ) + \log p( \vbeta \mid \sigma^2, b_{1:G}) \right ] \right] \\
		&\propto \exp \left[  \bE_q \left [ -\tfrac{n}{2} \log (\sigma^2) - \tfrac{1}{2 \sigma^2} \lVert \vy - \mX\vbeta \rVert^2 - \tfrac{1}{2 \sigma^2 \tau} \vbeta^\top \widetilde{\mD} (\vb) \vbeta\right ] \right] \\
		&\propto \exp \left[  \bE_q \left [ - \tfrac{1}{2} \vbeta^\top \left ( \mX^\top\mX/ \sigma^2 +  \tfrac{1}{\sigma^2 \tau}\widetilde{\mD} (\vb) \right ) \vbeta+ 2 \vy^\top \mX\vbeta/ \sigma^2 \right ] \right ] \\
		&\propto \exp \left[   - \tfrac{m_{1/\sigma^2}}{2} \vbeta^\top \left (  \mX^\top \mX  + \mM_\tau \right ) \vbeta+ 2 m_{1/\sigma^2} \vy^\top \mX\vbeta  \right ]
	\end{align*}
	Hence, $q(\vbeta) = \text{N} ( \vmu_\vbeta, \mSigma_\vbeta )$, where
	\begin{align*}
		\mSigma_\vbeta &= \frac{1}{m_{1/\sigma^2}} \left (  \mX^\top \mX  +  \mM_\tau \right )^{-1} \\
		\vmu_\vbeta &=  \left (  \mX^\top \mX  +  \mM_\tau \right )^{-1} \mX^\top \vy
	\end{align*}
	and $\mM_\tau = \frac{1}{\tau} \text{dg}\{ m_{b_1} \vone_{p_1}, \ldots, m_{b_G} \vone_{p_G}  \}$.
}

\subsection{Derivation for $q(\sigma^2)$}
{\scriptsize
	\begin{align*}
		&q(\sigma^2) \\ 
		&\propto \exp \left[  \bE_q \left [ \log p( \vy \mid \vbeta, \sigma^2 ) + \log p( \vbeta \mid \sigma^2, b_{1:G}) + \log p(\sigma^2) \right ] \right] \\
		&\propto \exp \left[  \bE_q \left [ -\tfrac{n}{2} \log (\sigma^2) - \tfrac{1}{2 \sigma^2} \lVert \vy - \mX \vbeta \rVert^2 - \tfrac{p}{2} \log (\sigma^2) - \tfrac{1}{2 \sigma^2 \tau} \vbeta^\top \widetilde{\mD} (\vb) \vbeta- (r+1) \log(\sigma^2) - s/\sigma^2 \right ] \right] \\
		&\propto \exp \left[   - \left ( r + \tfrac{n + p}{2} + 1 \right ) \log (\sigma^2) - (1/\sigma^2) \left \{ s + \tfrac{1}{2} \left \{  \bE\lVert \vy - \mX \vbeta \rVert^2  +\bE \left (\sum_{g=1}^G (b_g/\tau) \lVert \vbeta_g \rVert^2 \right ) \right \} \right \}  \right] \\
		&\propto \exp \left[   - \left ( r + \tfrac{n + p}{2} + 1 \right ) \log (\sigma^2) - (1/\sigma^2) \left \{ s + \tfrac{1}{2} \left \{  \lVert \vy - \mX \vmu_{\vbeta} \rVert^2 + \tr( \mX^\top \mX \mSigma_\vbeta )  +\bE (\sum_{g=1}^G (b_g/\tau) \lvert \vbeta_g \rVert^2 ) \right \} \right \}  \right] \\
		&\propto \exp \left[  - (1/\sigma^2) \left \{ s + \tfrac{1}{2} \left \{  \lVert \vy - \mX \vmu_{\vbeta} \rVert^2 + \tr \left \{ \left ( \mX^\top \mX + \mM_\tau \right ) \mSigma_\vbeta \right \} + \sum_{g=1}^G (m_{b_g}/\tau)  \lVert \vmu_{\vbeta_g} \rVert^2    \right \} \right \}  \right] \\
		&\times \exp \left \{ - \left ( r + \tfrac{n + p}{2} + 1 \right ) \log (\sigma^2) \right \} 
\end{align*}}
Hence, $q(\sigma^2) = \text{InvGa} ( r_{\sigma^2}, s_{\sigma^2} )$, where
\begin{align*}
	r_{\sigma^2} &= r + \frac{n + p}{2}\\
	s_{\sigma^2} &= s + \tfrac{1}{2} \left [  \lVert \vy - \mX \vmu_{\vbeta} \rVert^2 + \tr \left \{ \left ( \mX^\top \mX + \mM_\tau \right ) \mSigma_\vbeta \right \} + \sum_{g=1}^G (m_{b_g}/\tau)  \lVert \vmu_{\vbeta_g} \rVert^2    \right ]
\end{align*}

\subsection{Derivation for $q(c_g)$}
\begin{align*}
	q(c_g) &\propto \exp \left[  \bE_q \left [ \log p( b_g \mid c_g ) + \log p(  c_g) \right ] \right] \\
	&\propto \exp \left[  \bE_q \left [ \tfrac{1}{2} \log (c_g) - c_g b_g - \tfrac{1}{2} \log(c_g) - c_g \right ] \right] \\
	&\propto \exp \left[  - (m_{b_g} + 1) c_g \right]
\end{align*}
Hence, $q(c_g) = \text{Exp} ( s_{c_g} )$, where $s_{c_g} = m_{b_g} + 1$.

\subsection{Derivation for $q(b_g)$}
\begin{align*}
	q(b_g) &\propto \exp \left[  \bE_q \left [ \log p(\vbeta_g \mid b_g, \sigma^2) +  \log p( b_g \mid c_g )  \right ] \right]  \\
	&\propto \exp \left[  \bE_q \left [ - \tfrac{p_g}{2} \log(\sigma^2 \tau/b_g) - \frac{b_g \lVert \vbeta_g \rVert^2}{2 \sigma^2 \tau} - \tfrac{1}{2} \log(b_g) - c_g b_g   \right ] \right]  \\
	&\propto \exp \left[  (\tfrac{p_g + 1}{2} - 1) \log (b_g) - b_g \left \{ m_{c_g} + m_{1/\sigma^2} \frac{\bE \lVert \vbeta_g \rVert^2}{2 \tau}  \right \} \right] \\
	&\propto \exp \left[  (\tfrac{p_g + 1}{2} - 1) \log (b_g) - b_g \left \{ m_{c_g} + m_{1/\sigma^2} \frac{  \lVert \vmu_{\vbeta_g} \rVert^2 + \tr( \mSigma_{\vbeta_g}) }{2 \tau}  \right \} \right]
\end{align*}
Hence, $q(b_g) = \text{Ga} ( r_{b_g}, s_{b_g} )$, where $r_{b_g} = (p_g + 1)/2$ and
$$
s_{b_g} = m_{c_g} + m_{1/\sigma^2} \frac{  \lVert \vmu_{\vbeta_g} \rVert^2 + \tr( \mSigma_{\vbeta_g}) }{2 \tau}.
$$
\subsection{Derivation for ELBO}
{\small
	\begin{align*}
		&\text{ELBO}(\tau) \\
		&= \bE_q \left [ \log p_\tau ( \vy, \vbeta, \sigma^2, \mathbf{b}, \mathbf{c} ) \right ] - \bE_q \left [ \log q ( \vbeta, \sigma^2, \mathbf{b}, \mathbf{c} ) \right ] \\
		&= \bE \left [ \log \phi_n(\vy; \mX \vbeta, \sigma^2 \mI) \right ] + \bE_q \left [ \log \phi_p(\vbeta; 0, \sigma^2 \tau  \text{dg} \{ b_1 \vone, \ldots, b_G \vone \}^{-1} ) \right ] + \bE [ \log p(\sigma^2) ] + \sum_{g=1}^G  \bE_q \log p(b_g \mid c_g) \\
		&+  \sum_{g=1}^G  \bE_q \log p(c_g) - \bE_q \left [ \log q ( \vbeta, \sigma^2, \mathbf{b}, \mathbf{c} ) \right ] \\
		&= - \frac{n}{2} \bE \log (\sigma^2) - \frac{m_{1/\sigma^2}}{2}  \bE \lVert \vy - \mX \vbeta \rVert^2 - \frac{p}{2} \bE \log (\sigma^2)  - \frac{p}{2} \log (\tau) + \sum_{g=1}^G \frac{p_g}{2} \bE \log (b_g)  - \frac{  m_{1/\sigma^2} }{2 \tau} \sum_{g=1}^G   m_{b_g} \bE \lVert \vbeta_g \rVert^2\\
		&+ r \log(s) - \log \Gamma(r) - (r+1) \bE \log(\sigma^2) - s m_{1/\sigma^2} + \sum_{g=1}^G \left \{ \tfrac{1}{2} \bE \log (c_g) - \log \Gamma(\tfrac{1}{2}) + (\tfrac{1}{2} - 1) \bE \log (b_g) - m_{b_g} m_{c_g} \right \} \\
		&+ \sum_{g=1}^G \left \{ - \log \Gamma (\tfrac{1}{2}) - \tfrac{1}{2} \bE \log (c_g) - m_{c_g} \right \} - \frac{n+p}{2} \log(2\pi) - \bE_q \left [ \log q ( \vbeta, \sigma^2, \mathbf{b}, \mathbf{c} ) \right ] \\
		&= - \frac{n+p}{2} \log(2\pi) - \frac{p}{2} \log(\tau) - \frac{m_{1/\sigma^2}}{2 \tau} \sum_{g=1}^G m_{b_g} \bE \lVert \vbeta_g \rVert^2 - \left ( \tfrac{n+p}{2} + r + 1 \right ) \bE \log (\sigma^2) - m_{1/\sigma^2} \left \{ s + \tfrac{1}{2} \bE \lVert \vy - \mX \vbeta \rVert^2 \right \} \\
		& - 2G \log \Gamma(\tfrac{1}{2}) + \sum_{g=1}^G \left \{ \tfrac{1}{2} (p_g  - 1)    \bE \log (b_g) - (m_{b_g} + 1) m_{c_g}  \right \} + r\log(s) - \log\Gamma(r) - \bE_q \left [ \log q ( \vbeta, \sigma^2, \mathbf{b}, \mathbf{c} ) \right ] \\
		&= - \frac{n+p}{2} \log(2\pi) - \frac{p}{2} \log(\tau) - \frac{m_{1/\sigma^2}}{2 \tau} \sum_{g=1}^G m_{b_g} \bE \lVert \vbeta_g \rVert^2 - \left ( \tfrac{n+p}{2} + r + 1 \right ) \bE \log (\sigma^2) - m_{1/\sigma^2} \left \{ s + \tfrac{1}{2} \bE \lVert \vy - \mX \vbeta \rVert^2 \right \} \\
		& - 2G \log \Gamma(\tfrac{1}{2}) + \sum_{g=1}^G \left \{ \tfrac{1}{2} (p_g  - 1)    \bE \log (b_g) - (m_{b_g} + 1) m_{c_g}  \right \} + r\log(s) - \log\Gamma(r) + \tfrac{p}{2} \log (2\pi) + \tfrac{1}{2} \log \lvert \mSigma_\vbeta \rvert\\
		&+ \frac{p}{2} - r_{\sigma^2} \log( s_{\sigma^2} ) + \log \Gamma (r_{\sigma^2}) + (r_{\sigma^2} + 1) \bE \log (\sigma^2) + r_{\sigma^2} - \sum_{g=1}^G \Big \{ r_{b_g} \log(s_{b_g}) - \log \Gamma (r_{b_g}) \\
		&+ (r_{b_g} - 1) \bE \log (b_g)   - r_{b_g} + \log(s_{c_g}) - 1  \Big \} \\
		&= - \tfrac{n}{2} \log (2\pi)  - \tfrac{p}{2} \log(\tau) - m_{1/\sigma^2} \left [ \frac{1}{2 \tau} \sum_{g=1}^G m_{b_g} \bE \lVert \vbeta_g \rVert^2 + \tfrac{1}{2} \bE \lVert \vy - \mX \vbeta \rVert^2 + s  \right ] - 2G \log\Gamma(\tfrac{1}{2}) \\
		&+ \sum_{g=1}^G \left \{ \log \Gamma \{ (p_g + 1)/2 \} - \log (s_{c_g}) - \tfrac{1}{2} (p_g + 1) ( \log(s_{b_g}) - 1) -(m_{b_g}+1)m_{c_g}  \right \} + r \log(s) - \log \Gamma(r)  \\
		&+ \tfrac{p}{2} (  \log(2\pi) + 1) + \tfrac{1}{2} \log \lvert \mSigma_\vbeta \rvert - (r + \tfrac{n+p}{2}) \log(s_{\sigma^2}) + \log \Gamma\{ r + (n+p)/2 \} + r + \frac{n+p}{2} \\
		&= \text{const.} - \tfrac{p}{2} \log(\tau) - m_{1/\sigma^2} \left [ \frac{1}{2 \tau} \sum_{g=1}^G m_{b_g} \bE \lVert \vbeta_g \rVert^2 + \tfrac{1}{2} \bE \lVert \vy - \mX \vbeta \rVert^2 + s  \right ] + \tfrac{1}{2} \log \lvert \mSigma_\vbeta \rvert - (r + \tfrac{n+p}{2}) \log(s_{\sigma^2}) \\
		&- \sum_{g=1}^G \left \{ \log (s_{c_g}) + \tfrac{1}{2} (p_g + 1) ( \log(s_{b_g})) + \frac{m_{b_g}+1}{s_{c_g}} \right \} 
	\end{align*}
}
\newpage 

\subsection{Coordinate ascent update algorithm}
\begin{algorithm}[h]
	\caption{Block CAVI algorithm to compute variational parameters based on grouped horseshoe prior}
	\label{alg:CAVIhorseshoe}
		\KwSty{Initialize} $\vmu_\vbeta^{(0)} $, $\mSigma_{\vbeta}^{(0)}$,$\{ m_{b_g}^{(0)} \}_{g=1}^G$, $m_{1/\sigma^2}^{(0)}$, number of CAVI cycles $T$ \;
		\While{$\text{ELBO}^{(t)} - \text{ELBO}^{(t-1)}$ is large}{
			Compute $m_{b_g}^{(t)} \leftarrow (p_g + 1)\left \{\tfrac{2}{1 + m_{b_g}^{(t-1)}} + (\lVert \vmu_{\vbeta_g}^{(t-1)} \rVert^2 + \text{tr}(\mSigma_{\vbeta;gg}^{(t-1)}) )m_{1/\sigma^2}^{(t-1)} /\tau  \right \}^{-1} $\;
			Update $\mM_\tau^{(t)} \leftarrow \text{BlockDiag} (\{ m_{b_g}^{(t)} \mI_{p_g}  \}_{g=1}^{G} )/\tau$ \;
			Update $\vmu_{\vbeta}^{(t)} \leftarrow m_{1/\sigma^2}^{(t-1)} \mSigma_{\vbeta}^{(t-1)} \mX^\top \vy $\;
			Update $s_{\sigma^2} ^{(t)} \leftarrow s + \tfrac{1}{2} \left [  \sum_{g=1}^G \tfrac{m_{b_g}^{(t-1)}}{\tau} \left \{  \lVert \vmu_{\vbeta;g}^{(t-1)} \rVert^2 + \tr( \mSigma_{\vbeta;gg}^{(t-1)} )  \right \} + \tr( \mX^\top \mX \mSigma_{\vbeta}^{(t-1)}  ) + \lVert \vy - \mX \vmu_{\vbeta}^{(t-1)} \rVert^2 \right ]$\;
			Compute $m_{1/\sigma^2}^{(t)} \leftarrow \{r + (n+\sum_g p_g)/2 \} / s_{\sigma^2} ^{(t)} $\;
			Update $\mSigma_{\vbeta}^{(t)} \leftarrow (1/m_{1/\sigma^2}^{(t)}) \left \{  \mX^\top \mX +  \mM_\tau^{(t)} \right \}^{-1}$ \;
		}
		\KwSty{Set}  $\widehat{\vbeta} = \vmu_{\vbeta}^{(T)}$ and $\widehat{\mSigma} = \mSigma_{\vbeta}^{(T)}$. 
\end{algorithm}

\section{Additional details for the NHANES data example}

The predictors being considered in the NHANES example are: 
\begin{itemize}
	\item 18 organic pollutants, which include
	\begin{itemize}
		\item 11 polychlorinated biphenyls (PCBs): PCB74, PCB99, PCB118, PCB138, PCB153, PCB170, PCB180, PCB187, PCB194, PCB169, PCB126
		\item Three Dioxins, referred to as Dioxin1, Dioxin2, Dioxin3 
		\item Four Furans, referred to as Furan1, Furan2, Furan3, Furan4
	\end{itemize}
	
	\item 13 demographic variables: age, age squared, gender, BMI (categorical variable with 3 levels), education status (categorical variable with 4 levels), race (categorical variable with 4 levels), white blood cell count, lymphocyte count, monocyte count, cotinine level, basophil count, eosinophil count, and neutrophil count. 
\end{itemize}

\subsection{Selection rate of all covariates in 100 iterations of 10-fold cross validation}
\includepdf[pages=-]{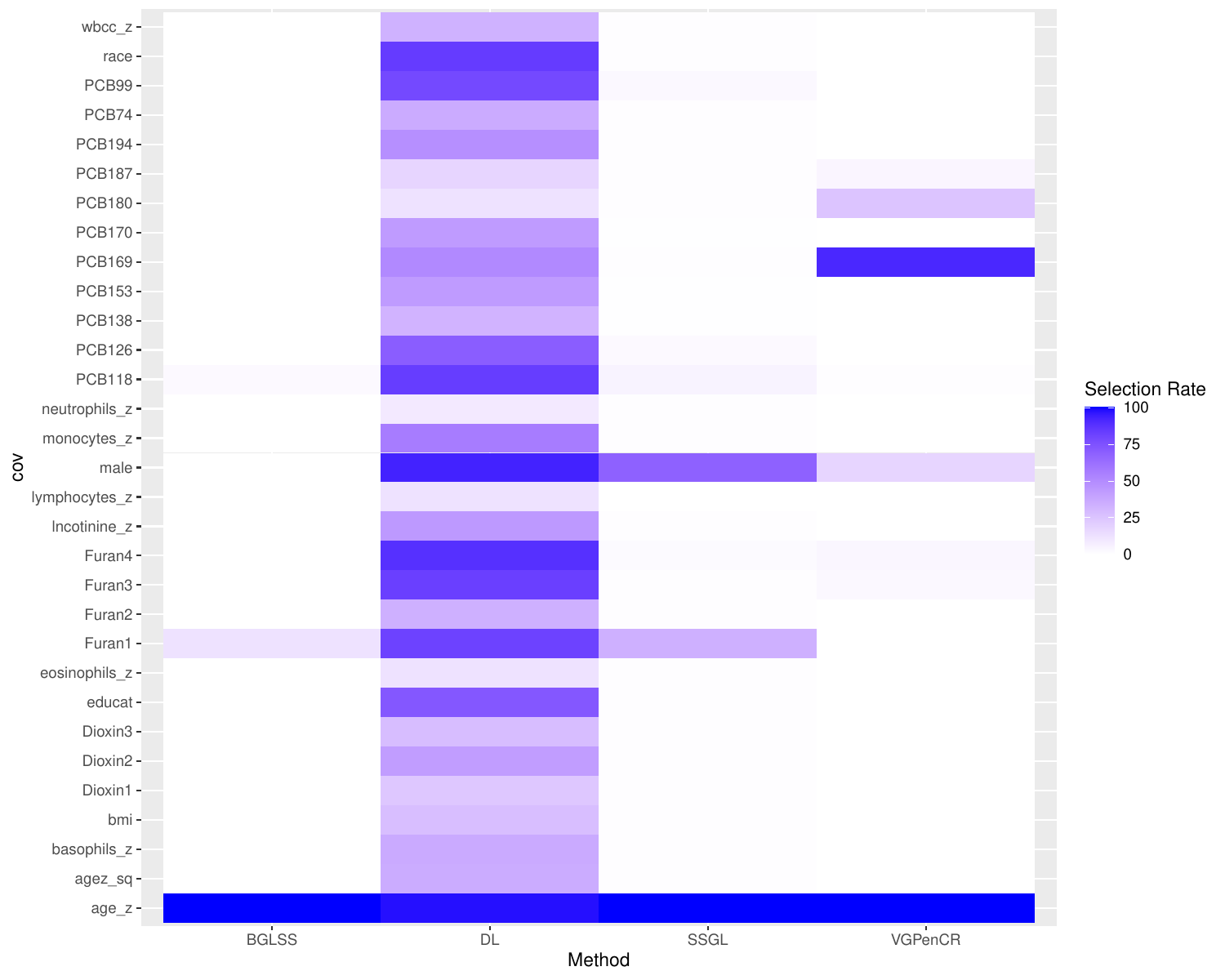}



\end{document}